\documentclass[10pt,journal,compsoc]{IEEEtran}
 \usepackage{multicol, blindtext} 
 \usepackage{hyperref}
 \usepackage{multirow}  
 \usepackage{graphicx,caption,subcaption}
 \usepackage{soul}
 \usepackage[utf8]{inputenc}
 \DeclareCaptionLabelFormat{AppendixTables}{Appendix \Alph{table}}
 \usepackage{afterpage}  
 \usepackage{longtable}  
 \usepackage{tcolorbox}
 \usepackage{xcolor,colortbl}
\usepackage[ruled]{algorithm2e} 

\newcommand{\inserted}[1]{\textcolor{black}{#1}}
\renewcommand{\small}{\fontsize{6pt}{7pt}\selectfont}

\ifCLASSOPTIONcompsoc
  \usepackage[noadjust]{cite}
\else
  \usepackage{cite}
\fi
\ifCLASSINFOpdf
\else
\fi
\usepackage{ragged2e}
\usepackage{url}
\renewcommand{\footnotesize}{\fontsize{6pt}{7pt}\selectfont}

\begin{document}
%
\title{Reliability and Robustness analysis of Machine Learning based Phishing URL Detectors}

\author{\IEEEauthorblockN{Bushra Sabir\IEEEauthorrefmark{1,2,3},
M. Ali Babar\IEEEauthorrefmark{1,2},
Raj Gaire\IEEEauthorrefmark{3}, 
Alsharif Abuadbba \IEEEauthorrefmark{3}, 
}
\IEEEauthorblockA{\IEEEauthorrefmark{1} School of Computer Science, The University of Adelaide}
\IEEEauthorblockA{\IEEEauthorrefmark{2} CREST - The Centre for Research on Engineering Software Technologies}
\IEEEauthorblockA{\IEEEauthorrefmark{3} CSIROs Data61}
}

\newcommand{\sharif}[1]{\textcolor{teal}{[Sharif: #1]}}

\IEEEtitleabstractindextext{%
\begin{abstract}
 \small \justifying
ML-based Phishing URL (MLPU) detectors serve as the first level of defence to protect users and organisations from being victims of phishing attacks.
Lately, few studies have launched successful adversarial attacks against specific MLPU detectors raising questions on their practical reliability and usage.
Nevertheless, the robustness of these systems has not been extensively investigated. Therefore, the security vulnerabilities of these systems, in general, remain primarily unknown that calls for testing the robustness of these systems.
In this article, we have proposed a methodology to investigate the reliability and robustness of 50 representative state-of-the-art MLPU models. Firstly, we have proposed a cost-effective Adversarial URL generator URLBUG that created an Adversarial URL dataset ($Adv_\mathrm{data}$) . Subsequently, we reproduced  50 MLPU (traditional ML and Deep learning) systems and recorded their baseline performance. 
Lastly, we tested the considered MLPU systems on $Adv_\mathrm{data}$ and analyzed their robustness and reliability using box plots and heat maps.
Our results showed that the generated adversarial URLs have valid syntax and can be registered at a median annual price of \$11.99, and out of 13\% of the already registered adversarial URLs, 63.94\% were used for malicious purposes. Moreover, the considered MLPU models Matthew Correlation Coefficient (MCC) dropped from median 0.92 to 0.02 when tested against $Adv_\mathrm{data}$, indicating that the baseline MLPU models are unreliable in their current form. Further, our findings identified several security vulnerabilities of these systems and provided future directions for researchers to design dependable and secure MLPU systems.
\end{abstract}

}

\maketitle

\IEEEdisplaynontitleabstractindextext

%
\IEEEpeerreviewmaketitle

\IEEEraisesectionheading{\section{Introduction}
\label{introduction}}
\IEEEPARstart{P}{hishing} attacks are a critical security threat, and their success leads to financial and reputational damage to a system and its users \cite{sabir2020machine}.
For instance, Xoom corporation, a money transfer company, accidentally transferred \$30.8m of corporate cash to fraudulent overseas accounts due to a phishing attack.
While Xoom lost a large sum of money, the company stock also dipped by 14\%\footnote{\url{https://bit.ly/3z9SCZ2}}$^,$\footnote{\url{https://bit.ly/3qCFOqa}}.
Moreover, a successful phishing attack can open the gateway for more sophisticated cyber attacks such as Advanced Persistent Threat (APT) \cite{daly2009advanced}, ransomware \cite{thomas2018individual}, and data breaches \cite{sabir2020machine}. Verizon \cite{Verizon2021} latest report pointed out that in 2021, 80\% of data breaches happened as a result of successful phishing attacks. 
Therefore, \textbf{timely} and \textbf{proactive} detection of phishing attacks is essential to secure the cyber-infrastructure.
\par
\inserted{
To this end, several efforts have been made by the industry and academia to prevent people from being a victim of phishing attacks.
Existing phishing detection methods are classified in literature \cite{varshney2016survey,dou2017systematization,sahoo2017malicious} into two main categories: list-based and Machine Learning (ML) based approaches. 
\textit{List-based} solutions, search a given URL in a list of known URLs (blacklists: list of phishing URLs or whitelists: list of legitimate URLs) to detect its phish and legit nature \cite{virustotal, Overview8:online,isitPhis11:online, Kaspersk51:online, IsThisWe33:online, MalwareU22:online, PhishTan46:online}. These are effective for detecting known phishing attacks but are insufficient to detect novel (zero-day) attacks.
Moreover, with billions of URLs being registered every day \cite{ma2009beyond}, comparing a URL and constantly updating these lists is infeasible at scale.
On the other hand, \textit{ML-based} approaches can detect zero-day attacks \cite{sahoo2017malicious, Shirazi2018KnowThyDomainName}.
 These approaches analyse different aspects of a web resource to detect its phishing or legit nature and are classified into three main types:
Web Content-based (WCB), Visual Similarity-based (VSB).
and URL-based (MLPU)}
\par
\inserted{
WCB detectors obtain features from the web resource source code i.e., HTML, extensible mark-up language (XML), JavaScript (JS), and CSS for performing classification \cite{sahoo2017malicious,Shirazi2018KnowThyDomainName,li2019stacking,lee2020building}.
Most anti-phishing entities such as Virus-Total vendors~\cite{virustotal} use them to detect phishing attacks \cite{crackingclassifier}. 
These solutions perform an in-depth analysis of a web resource that on one side enables them to capture complex phishing patterns but conversely results in computational burden and inefficiency in real-time.
Moreover, they need domain experts (to extract useful features) and pose a high-security risk \cite{sahoo2017malicious} as parsing a web resource can lead to the execution of the malicious code \cite{bahnsen2018deepphish}.}
\par
\inserted{Recently, VSB approaches have gained popularity \cite{jain2017phishing}. These approaches compare screenshots, and logos of well-known brands and train Machine learning-based algorithms to compare them with phishing web pages automatically  \cite{lin2021phishpedia,abdelnabi2020visualphishnet,liu2022inferring}. The web pages having a high correlation with well-known brands are considered to be phishing web pages.
Similar to WCB, VSB solutions are computationally costly, pose of high-security risk and are inefficient in real-time \cite{sahoo2017malicious,jain2017phishing}. Moreover, these solutions can only protect the top brands as maintaining and comparing billions of screenshots of millions of legitimate e-commerce or banking web pages is infeasible. For example, the latest Usenix Seurity'22 work named ``PhishIntention'' \cite{liu2022inferring} could only protect 277 brands maximum, whereas phishpedia \cite{lin2021phishpedia} collected logos from only 188 brands and another seminal work VisualPhishNet \cite{abdelnabi2020visualphishnet} could only safeguard 155 brands. 
Moreover, most of the VSB solutions are sensitive to rotation and layout (\cite{jain2017phishing} provides more details) and generate FPR e.g., when a site put a logo of Instagram \cite{liu2022inferring}}
\par
\inserted{
Lastly, \textit{MLPU} approaches to utilize the address (URL) of a web resource to detect whether it resembles phishing or legit URLs. They are the most common \cite{sabir2020machine}, light-weight \cite{bahnsen2018deepphish}  and safe solutions \cite{sahoo2017malicious} (don't require parsing or visiting a web resource) to detect phishing. That is why, they serve as first-level of defence against phishing attacks and are deployed in client-side anti-phishing browser's \cite{bahnsen2018deepphish} and email server plugin \footnote{\url{https://bit.ly/3Cwn80t}}  \cite{verma2017s,bahnsen2017classifying,urlnet}.  For instance, the Google Chrome extension Sharkcop uses MLPU to warn users about a phishing URL \footnote{\url{https://bit.ly/37wuvqs}}. 
These detectors have shown good accuracy and effectiveness in detecting zero-day phishing URLs \cite{das2019sok} and are considered as an \textbf{effective}, \textbf{timely} and \textbf{scalable} defense against phishing attacks \cite{bahnsen2018deepphish,tajaddodianfar2020texception,haynes2021lightweight,varshney2016survey,sahoo2017malicious,dou2017systematization,Aldwairi2017DrivebyDownload}. }

\inserted{
Nevertheless, ML-based solutions are more effective than list-based phishing defence strategies but recent studies \cite{szegedy2013intriguing, PapernotMGJCS2016BlackboxAttacks,biggio2018wild, Fass2019HideNoSeek} have unveiled that ML solutions, yield unreliable results at test time when subjected to adversarial examples.
Adversarial examples are artificially synthesised input examples on which ML models tend to misclassify the output classes, e.g., classify malicious samples as benign. 
Cybercriminals can bypass ML models by subjecting them to adversarial examples which can hinder their practical capabilities. For instance, in 2019 the first vulnerability against ML-based commercial email protection system named Proofpoint\footnote{ \url{https://www.proofpoint.com/us/security/security-advisories/pfpt-sn-2020-0001}} was reported in National Vulnerability database\footnote{ \url{https://nvd.nist.gov/vuln/detail/CVE-2019-20634}}.
Attackers exploited a vulnerability by copying the underlying ML model and then launching an attack by crafting emails that allowed the malicious (phishing) email to bypass the Proofpoint system. 
Such an attack can increase the chances of unwitting users being victims of phishing attacks\cite{Chiew2018}. 
Therefore, it is necessary to study 
efficient and practical evasion techniques against ML systems to unveil their security vulnerabilities.}

This work investigated the robustness of ML-based Phishing URL (MLPU) detectors to unveil their security vulnerabilities.
\subsection{\inserted{Motivation}}
\label{motive}
\inserted{The motivation behind examining the robustness of MLPU detectors is as follows:}\par
\inserted{\textbf{(1) Constraints of other ML-based approaches.}
Although, all the above-mentioned ML approaches are effective in detecting phishing attacks in practice they have three challenges in comparison to the URL-based approach.  
\textit{ (i) Dataset Collection}: A large labelled dataset is required for training the models, as phishing websites are short-lived it might not be easy to collect data from them in a timely fashion. Furthermore, with emerging threats the dataset becomes obsolete \cite{liu2022inferring} very quickly and therefore, an up-to-date labelled dataset is required to retrain these models over time \cite{lin2021phishpedia} which hinders the {proactiveness} and {scalability}. This challenge is more significant for WCB and VSB approaches as they require a web resource to be online to collect useful features or screenshots whereas for URL-based methods if the website becomes offline, the URL can still be used for capturing phishing patterns.
(ii) \textit{Timely Detection:} A recent study \cite{oest2020sunrise} showed that an average phishing attack may last for 21 hours 
while their detection by well-known anti-phishing entities occurs on average 9 hours after the first victim visit. Further, seven hours elapse occurs before these entities block them.
This delay is sufficient for an attacker to execute a fraud, and more timely measures are required to mitigate such attacks. However, WCB and VSB solutions are both computation and resource extensively which hinders their \textbf{timeliness}.
(iii) Scalability: VSB solutions can only protect the top brands as maintaining and comparing billions of screenshots of millions of legitimate e-commerce or banking web pages is infeasible. For instance the latest Usenix Seurity'22 work named ``PhishIntention'' \cite{liu2022inferring} could only protect 277 brands maximum. For other domains, these approaches are not scalable. However, WSB and MLPU detectors capture the phishing patterns that are generalizable across other domains.}
\par\inserted{\textbf{(2) Gaps in Literature.}
 Several efforts have been done by researcher \cite{abuadbba2022towards,crackingclassifier,AdversarialSampling,almashor2021characterizing,song2021advanced,bac2021pwdgan,wong2022phishclone} to identify the vulnerabilities in ML-based phishing classifiers. However, most of them are limited to WCB classifiers. For example, authors in the study \cite{abuadbba2022towards} pointed out five obfuscation methods used by attackers to bypass these systems, these include the usage of benign web services to camouflage phishing pages, enhance the similarity between the HTML structure of benign and phishing pages and hide the HTML content behind scripts. 
Conjointly, VSB classifiers use computer vision methods to detect a phishing website and there has been a significant amount of work on identifying the vulnerabilities of these types of models \cite{biggio2018wild}. 
}\par\inserted{
However, MLPU models are not extensively studied in terms of evasion and their vulnerabilities remain primarily unknown. Although, few preliminary studies \cite{DeepPhis17, AlEroud2020GAN} have demonstrated that specific MLPU models can be fooled by generating adversarial examples using Deep Neural Networks (DNN).
These studies have several limitations. (i) Lack of comprehensive evaluation: The studies generate adversarial examples to fool a specific MLPU model.
Therefore, it is hard to comprehend their impact and generalizability across other MLPU models.
(ii) Computationally expensive: they require training DNN to generate adversarial examples and need the MLPU system's feedback to select the seed URLs (URLs used to create new adversarial examples).
(iii) Realizability Evaluation: they do not evaluate the generated Adversarial Examples (AEs) against lexical structure validity (whether the generated URL follows RFC standard \cite{RFC3986U96:online}), computational effort, annual registration cost and deception (whether the generated URL obfuscate the identity of existing URLs or not). }
\par\inserted{\textbf{(3) Approaches from other domains are not transferable.}
Evasion techniques developed for the computer vision domain such as gradient-based adversarial attacks will not directly translate to discrete text space \cite{wang2022measure}. Similarly, attacks on other phishing and Natural Language Processing (NLP) \cite{evans2021raider,jin2019bert,kashapov2022email,sabir2021reinforcebug,shmalko2022profiler} models are not directly applicable to MLPU detectors because of the well-constraint nature of URL formulated by RFC 3986 standard \cite{RFC3986U96:online}. For example, the authors in a recent study \cite{song2021advanced} systematically studied the robustness of WCB detectors by launching three mutation attacks. However, they pointed out that their method did not perturb the URL due to the difficulty of automatically mutating URLs that can retain similarity to the victim's URLs. 
Similarly, approaches from the NLP domain cannot be directly applied. For instance, a study \cite{jin2019bert} used the synonym substitution technique to generate realistic AEs and used semantic and grammatical similarity to evaluate the realizability of generated AEs automatically. However, this method cannot be applied to URLs. Unlike text, a URL is a \textit{sequence of characters} instead of meaningful English words, and synonym substitution is not possible in most cases. Despite that char-level attacks such as DeepWordBug \cite{gao2018black} can be used to augment data. We have already considered valid perturbations (according to URL format RFC3986) from DeepWordBug while designing our URLBUG method.   }

To fill the aforementioned crucial gaps, our study has comprehensively and systematically evaluated the robustness of MLPU models.
\subsection{Our Approach}
In this article, we have comprehensively and systematically evaluated the robustness of 50 representative MLPU models.
We have considered both traditional ML and Deep learning models in our benchmark.
Firstly, we have devised a simple, generic and fast method called URLBug to generate Adversarial URLs (AUs) by formulating three types of adversaries that obfuscate three parts of the URL, i.e., domain, path, and Top Level Domain (TLD). We automated 16 different URL obfuscation methods to generate these AUs.
Secondly, we reproduced 50 baseline MLPU models that were frequently reported in the literature \cite{das2019sok}. After that, we analysed the AUs for their realizability by investigating their computational, registration cost and deceptive.
We then analysed the robustness of 50 baseline MLPU models against the generated AUs and identified the most robust features, classifiers, and consequently reliable MLPU models. Furthermore, we highlighted and discussed the limitation and vulnerabilities in the baseline MLPU models and provided recommendations to assist future research.
 The \textbf{novel contributions} of our work are as follows: 
\begin{itemize}
 \item Cost-effective adversarial URL generator URLBUG 
  \item An Adversarial URL (AU) dataset to test the robustness of MLPU models and assist the user's training. 
 \item A comprehensive experimental study to test the robustness of 50 state-of-the-art MLPU systems against the adversarial dataset.
\item Identification of vulnerabilities in the considered MLPU systems.
\item  Investigation of the impact of adversarial defence methods Adversarial Training and Ensemble on the MLPU systems.
\item Reproducible code for AU dataset generation and 50 trained MLPU models.
\end{itemize}
The remainder of this paper is organised as follows.
Section~\ref{ourapproach} describes our methodology and detailed experimental design. 
We report our results in section~\ref{results} while section~\ref{discussion} discusses the security vulnerabilities in baseline MLPU models and their evaluation challenges. and finally, section~\ref{conclude} concludes the paper.

\begin{table*}[hbt!]
\caption{Baseline MLPU systems [Empty cells depict the metrics is not reported by the studies]}
\label{stateofart}
\centering
\resizebox{0.8\textwidth}{!}{\begin{tabular}{l|c|c|c|c|c|c|c}
\hline
\textbf{MLPU Type} &\textbf{Study} &\textbf{Features Type} &\textbf{Classifier} & \textbf{Accuracy} &\textbf{Precision} &\textbf{Recall} & \textbf{AUC}\\
\hline
\multirow{16}{2em}{T$-$MLPU}
&\cite{kolari2006svms}  & BoW & SVM & &$>98$\%&$>94$\%&\\
 \cline{2-8}
 &\multirow{2}{*}{\cite{ma2009beyond}}& BoW of URL parts,  & SVM, & \multirow{2}{*}{ $>99$\%} &&&\\
 && External &  &&&&\\ 
 \cline{2-8}
 &\cite{Feroz2014Examination} & Bigram & LR & $>94$\%&&&\\
 \cline{2-8}
 &\cite{Darling2015lexicalmaliciousURL} & Character n-grams & DT &&&&$>99$\%\\
  \cline{2-8}
&\multirow{2}{*}{\cite{Vanhoenshoven2016MaliciousURLs}} & BoW of URL parts, & RFT, SVM & \multirow{2}{*} {$>95$\%} & \multirow{2}{*}{$>95$\%} & \multirow{2}{*}{$>95$\%}&\\
&&External &  KNN, DT&&&&\\
 \cline{2-8}
&\cite{Mamun2016LexicalMaliciousURL}  & Basic Lexical & KNN, DT, RFT & & $>97$\% &$>97$\% & \\ 
 \cline{2-8}
&\cite{verma2017s}  
& Character n-grams & RFT & $>99$\% &&&\\
 \cline{2-8}
&\multirow{2}{*}{\cite{jain2018phishsafe,hong2020phishing}}& Basic Lexical+ & \multirow{2}{*}{SVM} &\multirow{2}{*}{ $>91$\%} &&&\\
&&External&&&&&\\
 \cline{2-8}
&\multirow{2}{*}{\cite{Liu2018find}}  & \multirow{2}{*}{Bigram}  & XGB & \multirow{2}{*}{$>95$\%}& \multirow{2}{*}{$>95$\%} & \multirow{2}{*}{$>93$\%}&\\
&&&LGBM&&&&\\
  \cline{2-8}
 &\cite{patgiri2019empirical} & BoW  & RFT & $>90$\% &&&\\
 \hline
 \multirow{6}{5em}{DNN$-$MLPU}
  &\cite{Saxe2017Expose}  & Character vector & CNN &&&&$>99$\%\\
   \cline{2-8}
 &\cite{bahnsen2017classifying}   & Character vector & LSTM & $>98$\% &&&\\
  \cline{2-8}
 &\multirow{2}{*}{\cite{urlnet}}  & Word and & \multirow{2}{*}{CNN}&&&&\multirow{2}{*}{$>99$\%}\\
 &&Character vector &&&&&\\
  \cline{2-8}
&\cite{zhao2018classifying}   & Basic Lexical & GRU & $>98$\%&&&\\
 \cline{2-8}
&\cite{shivangi2018chrome}  & Character vector & LSTM  & $>96\%$ &&&\\
\cline{2-8}
&\cite{afzal2021urldeepdetect}  & Word vector & LSTM  & $>98\%$ & $>99\%$& $>97\%$&\\
\hline
\end{tabular}}
\end{table*}

\section{Robustness Analysis Methodology and Experimental Setup}
\label{ourapproach}
Our proposed methodology is shown in Figure~\ref{figure0}.
The goal of our study was to test the robustness and reliability of a pre-trained MLPU model $F$: $X \rightarrow Y$, which mapped from input space X to a class $Y \in {phish, benign}$. To achieve this goal, we proposed a framework called URLBUG that generated a set of Adversarial URLs $AU$ by simulating three types of adversaries: domain, path and TLD. These adversaries obfuscated the identity of a legitimate URL $x$.
URLBUG generated $AUs$ independent of $F$ and did not require any knowledge about $F$. 
After that, we reproduced 50 baseline MLPU models that represented the state-of-the-art (Table~\ref{stateofart}). 
Lastly, we evaluated the realizability of URLBUG and the adversarial robustness of MLPU models to answer the three research questions enlisted in Table~\ref{tab1}. 
The details of each step is described below:
\begin{figure}[!tb]
\centerline{\includegraphics[width=\columnwidth]{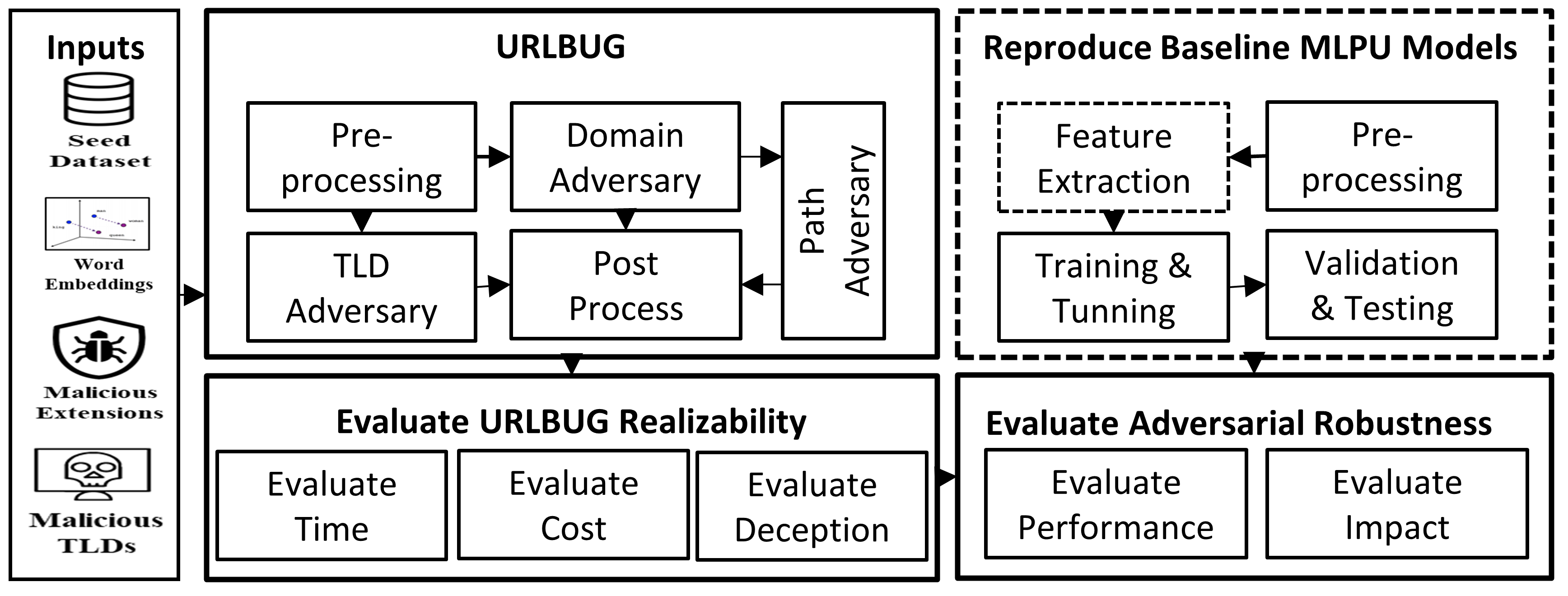}}
\caption{Overview of our methodology (The dotted line for depicts the optional modules)}
\label{figure0}
\end{figure}
\begin{table*}[tb!]
  \caption{Research Questions and their motivation}
  \label{tab1}
\begin{small}
\centering
\begin{tabular}{>{\raggedright\arraybackslash}m{5mm}|m{35mm}|m{50mm}|m{60mm}}
\hline
\multicolumn{1}{>{\centering\arraybackslash}m{5mm}|}{\textbf{ID}} &
\multicolumn{1}{>{\centering\arraybackslash}m{35mm}|}{\textbf{Research Questions}} 
    & \multicolumn{1}{>{\centering\arraybackslash}m{50mm}|}{\textbf{Sub-Research Questions}} 
    & \multicolumn{1}{>{\centering\arraybackslash}m{60mm}}{\textbf{Motivation}}\\
\hline
\multirow{3}{*}{\textbf{RQ1}}& Does URLBUG generate realistic $AUs$ ?& \textbf{RQ1.1)} How much computational effort is required to generate AUs against $x$?   \par         
\textbf{RQ1.2)} What is the annual registration price of a generate $AUs$?
\par         
\textbf{RQ1.3)} Are the generated $AUs$ deceptive?& 
Before evaluating the robustness of MLPU models, it is crucial to verify that URLBUG generates $AUs$ they pose a real-threat to MLPU models.\\
\hline
\multirow{2}{*}{\textbf{RQ2}}& Are the baseline MLPU systems robust against generated AUs ?&\textbf{RQ2.1)} What type of MLPU models (traditional versus deep learning based) are more robust? \par 
\textbf{RQ2.2)} Which classifiers are more robust? \par
\textbf{RQ2.3)} Which features are more robust?& 
Such an analysis can help developers select robust models, classifiers, and features for developing MLPU systems. Whilst it can also assist researchers in identifying security vulnerabilities in these systems. \\
\hline
\multirow{2}{*}{\textbf{RQ3}}& What is the impact of different adversaries and URL obfuscation techniques on the output of MLPU systems?&\textbf{ RQ3.1)} What type of adversarial URLs decrease the MLPU performance? \par 
\textbf{RQ3.2)} Which URL obfuscation techniques influence the performance of MLPU models?& This analysis can guide researchers to carry out more specific research towards developing robust MLPU systems that are resilient against potent adversaries, perturbation levels and obfuscation techniques.\\
\hline
\end{tabular}
\end{small}
\end{table*}
\begin{figure*}[htb]
\begingroup 
\csname @twocolumnfalse\endcsname
\noindent
\resizebox{.49\textwidth}{!}{%
\begin{minipage}{.7\textwidth}
\begin{algorithm}[H]
   \caption{Domain Adversaries ${DM_{\mathrm{adv}}}$}
   \label{domain}
   \SetKwProg{generate}{Function \emph{generate}}{}{end}
   {\bfseries Input:} $url_t$,$d$, $n$,${w_{\mathrm{emb}}}$, $x$\;
   {\bfseries Output:} ${D_{\mathrm{u}}}$\;
   Initialize ${D_{\mathrm{a}}}$=[] //list of updated $url_t$\;
  \textit{\textbf{Generate Char-level adversaries}}\;
  $D_a.append(replace(url_t(d),\textbf{char\_obfuscate}(d)))$\;
   \textit{\textbf{Generate Word-level adversaries}}\;
   \uIf{if $d $ in ${w_{\mathrm{emb}}}$.vocab}{
    $R_n = {w_{\mathrm{emb}}}$.similar-words($d$,$n$)\; 
  }
  \Else{
   $R_n =  $\textbf{SymSpell}($d$,$n$)\;
  }

   \ForAll{$r$ in $R_n$}{
    {\bfseries ${url_{\mathrm{sub}}}$}.append($replace(url_t, d=concat(d$,`.',$r)$)\;
    {\bfseries ${url_{\mathrm{part}}}$}.append($replace(url_t,d = concat(d$,`op',$r)$)\;
    }
  \If{($number of words(d)>1)$}{
   ${url_{\mathrm{swap}}}$.append($replace(url_t,d=$swapparts$(url_t,d)))$\;
  }
  {\bfseries ${url_{\mathrm{repeat}}}$}.append($replace(url_t,d=concat(d$,`-',$d)$)\;
  $D_a.append$(${url_{\mathrm{sub}}}$)\;
  $D_a.append$(${url_{\mathrm{part}}}$)\;
  $PT_{\mathrm{adv}}({url_{\mathrm{part}}})//$send to path adversary \;
  $D_a.append$(${url_{\mathrm{swap}}}$)\;
  $D_a.append({url_{\mathrm{repeat}}})$\;
  $D_u=$postprocess($D_a$,$x$)\;
  {\bfseries return:} ${D_{\mathrm{u}}}$\;

\end{algorithm}
\end{minipage}%
}
\hfill
\resizebox{.49\textwidth}{!}{%
\begin{minipage}{.7\textwidth}

\begin{algorithm}[H]
   \caption{Path Adversaries ${PT_{\mathrm{adv}}}$}
   \label{path}
   \SetKwProg{generate}{Function \emph{generate}}{}{end}
   {\bfseries Input:} $url_t$,$d$,$p$ $n$,${w_{\mathrm{emb}}}$, $x$,${lst_{\mathrm{exe}}}$ ,${url_{\mathrm{part}}}$\;
   {\bfseries Output:} ${P_{\mathrm{u}}}$\;
   Initialize ${P_{\mathrm{a}}}$=[] //list of updated $url_t$\;
   {${d_n}={url_{\mathrm{part}}}$}\;
    \ForAll{$i$ {\bfseries in} $d_n$}{
        \ForAll{$loc$ in path}{
       {$p_{\mathrm{dm}} = replace$($d$,loc,$p$)}\;
        { rand=random(0, len(${lst_{\mathrm{exe}}}$))}\;
        \uIf{if('ext' in $p$)  //ext denotes the extension}{
          {$p_{\mathrm{exe}}=replace(ext, p_{\mathrm{dm}},{lst_{\mathrm{exe}}}[rand])$}\;}
        \Else{
          {$p_{\mathrm{exe}}=concat(p_{\mathrm{dm}},{lst_{\mathrm{exe}}}[rand])$}\;
        }
       {$P_a$.append($replace(url_t,d=i,p=p_{\mathrm{dm}}$}))\;
       {$P_a$.append($replace(url_t,d=i,p=p_{\mathrm{exe}}$}))\;
        
        }}
        
        $P_u=$postprocess($P_a$,$x$)\;
        {\bfseries return:} ${P_{\mathrm{u}}}$\;

\end{algorithm}
\begin{algorithm}[H]
\caption{TLD Adversary ${TD_{\mathrm{adv}}}$}
\label{tld}
 \SetKwProg{generate}{Function \emph{generate}}{}{end}
  {\bfseries Input:}$url_t$, $t$, $x$, ${l_{\mathrm{tld}}}$\;
   {\bfseries Output:} ${T_{\mathrm{u}}}$\;
     Initialize ${T_{\mathrm{a}}}$=[] \;
  \ForAll{$mt$ {\bfseries in} ${l_{\mathrm{tld}}}$}{
  {$T_{\mathrm{a}}$.append(replace($url_t$,$t$=$mt$))}\;
 }
  $P_u=$postprocess($T_a$,$x$)\;
  {\bfseries return:} ${T_{\mathrm{u}}}$\;

\end{algorithm}
\end{minipage}%
}
\endgroup
\caption{Algorithms for generating Adversaries}
\end{figure*}
\begin{table*}[htb]
\caption{List of URL Obfuscation Techniques and Examples of Generated Adversaries}
\label{tab2}
\centering
\begin{small}
 \begin{tabular}{>{\raggedright\arraybackslash}m{10mm}|m{10mm}|m{20mm}|m{50mm}|m{50mm}}
\hline
\multicolumn{1}{>{\centering\arraybackslash}m{8mm}|}{\textbf{Type }} 
&\multicolumn{1}{>{\centering\arraybackslash}m{5mm}|}{\textbf{Level}} 
    & \multicolumn{1}{>{\centering\arraybackslash}m{15mm}|}{\textbf{Obfuscation Method}} 
    & \multicolumn{1}{>{\centering\arraybackslash}m{40mm}|}{\textbf{Description}} 
    & \multicolumn{1}{>{\centering\arraybackslash}m{30mm}}{\textbf{Target (store.steampowered.com/) }}\\
\hline
\multirow{12}{*}{Domain}& 
\multirow{10}{*}{Char} & Addition &Adding a character on start or end of a domain&	store.steampowered\textbf{\textcolor{purple}{a}}.com \\ 
\cline{3-5}
&& Insertion &Inserting a character in the domain&	store.stea\textbf{\textcolor{purple}{o}}mpowered.com \\ 
\cline{3-5}
&&BitSquatting&Replacing a character in domain with one bit different character.& store.steampowere\textbf{\textcolor{purple}{l}}.com\\
\cline{3-5}
&&Homoglyph& Replacing a character with visually similar ASCII character.	& store.steamp\textbf{\textcolor{purple}{0}}wered.com\\
\cline{3-5}
&&Omission&Omitting one character.	& store.s\textbf{\textcolor{purple}{ }}eampowered.com\\
\cline{3-5}
&&SubDomain& Separating the tokens in the domain with a `.`. &store.steam\textbf{\textcolor{purple}{.}}powered.com\\
\cline{3-5}
&&Hyphenation &Separating the characters in the domain with a hyphen.&store.st\textbf{\textcolor{purple}{-}}eampowered.com\\
\cline{3-5}
&&CharacterSwap& Swapping two consecutive characters in domain.	&store.steampow\textbf{\textcolor{purple}{i}}red.com\\
\cline{3-5}
&&Repetition&Repeat the previous character in the domain.&store.stea\textbf{\textcolor{purple}{a}}mpowered.com\\
\cline{3-5}
&&Transpose&Swap a subset of characters in the domain.&store.ste\textbf{\textcolor{purple}{mo}}powered.com\\
\cline{2-5}
&\multirow{4}{*}{Word} & 
Word SubDomain& Add semantically word at the end of the domain separated by `.`. &	store.steampowered.\textbf{\textcolor{purple}{ai-assisted}}.com\\
\cline{3-5}
&&Word Hyphenation&Add semantically related word at the start or end of domain with or without hyphen.& store.\textbf{\textcolor{purple}{ai-assisted-}}steampowered.com	\\
\cline{3-5}
&&Word Repetition&Concatenate the domain token with original domain.&	store.steampowered-\textbf{\textcolor{purple}{steampowered}}.com\\
\cline{3-5}
&&WordSwap&If domain has more than two tokens swap each token&	store.\textbf{\textcolor{purple}{poweredsteam}}.com/\\
\hline
\multirow{2}{*}{Path} &\multirow{2}{*}{Word} & 
PathDm	&Use domain adversary as domain and add the original domain in the path.& store.steampowered\textbf{\textcolor{purple}{-operated}}.com/\textbf{\textcolor{purple}{steampowered}}\\
\cline{3-5}
&&PathExe&Use the domain adversary as domain and add a malicious extension to the path of the URL.&
store.steampowered\textbf{\textcolor{purple}{-operated}}.com/\textbf{\textcolor{purple}{steampowered.exe}}\\
\hline
\multirow{1}{*}{TLD}&\multirow{1}{*}{Word} &TldReplace  &Replace the TLD of the URL with malicious TLD.&	store.steampowered.\textbf{\textcolor{purple}{in.rs}}\\
\hline
\end{tabular}
\end{small}
\end{table*}
\subsection{URLBUG: An Adversarial URL Generator}
\label{urlbug}
To generate $AUs$, we have proposed an adversarial generator called URLBUG. URLBUG created three adversaries targeting specific URL parts, i.e., domain, path, and TLD. These adversaries obfuscated the identity of URLs in the seed dataset $dseed$.
The seed dataset contained a set of legitimate URLs that can be potential phishing targets.
To generate the adversaries, we used the following steps:

\textbf{(i) Pre-processing.} Each URL $x$ in $dseed$ was decomposed into meaningful tokens 
$url_t$: IP address (ip), port, scheme, sub-level domain (SLD), domain ($d$), tld ($t$), path ($p$), extension ($exe$). Further, we decomposed each part into more meaningful tokens using "English word-based" tokeniser \cite{baziotis-pelekis-doulkeridis:2017:SemEval2}. For instance, in 'hxxps://www.credit-suisse.com/au/en/private-banking/contact-us.html', was segmented into following tokens: \{ip:(),port:(),scheme: (hxxps), sld: (www), domain: (credit, suissue), tld: (com), path:(au, en, private, banking, contact, us), exe: (html)\}
These tokens served as an input to the adversaries described below:

\par{\textbf{(ii) Domain Adversary.}} 
\label{dm}
Domain adversary added noise to the original domain of the seed URL $x$ by using the algorithm~\ref{domain} and returned a list of obfuscated domains.
It took five inputs: $url_t$, domain $d$, number of similar words $n$, pre-trained word embedding ${W_{\mathrm{emb}}}$,
and then perturbed $d$ at two levels of granularity: character (char) and word.
The char-level granularity produced syntactically similar domain names by altering their characters. In contrast, our word-level adversary targeted the domain name by concatenating it with a similar word.
Overall, domain adversary automated 13 types of URL obfuscation techniques. These methods, along with their level of granularity, description and an example, are mentioned in Table~\ref{tab2}. 
For constructing, subdomain (${url_{\mathrm{sub}}}$) and part (${url_{\mathrm{part}}}$) word-level adversaries, our method queried $d$ in ${W_{\mathrm{emb}}}$ 
to find semantically similar words in it. 
If $d$ was in the vocabulary then a list of $n$ semantically similar words to $d$ were retrieved from ${W_{\mathrm{emb}}}$.
The rationale behind using a similar word was to improve the intra-relatedness of terms to convey a semantically coherent meaning in the adversarial URL.
If $d$ was not in the ${W_{\mathrm{emb}}}$ then we used SymSpell library \cite{wolfgarb61:online} to get the $n$ syntactically similar words. 
This library provided spelling correction suggestions for unknown words and constructed a list of $n$ words with similar spellings. 
For example, given a domain name which is not a valid word as `adcb' (Abu Dhabi Commercial Bank), using SymSpell, we obtain a list of $n$ syntactical similar terms such as 'abd', 'act', 'ads', 'acc' etc. 
We denoted the list of these $n$ similar words by $R_n$. For each similar word $r$ in the list $R_n$, the algorithm concatenated it with $d$ using ‘.’ and op i.e. ('-' or '') to create ${url_{\mathrm{sub}}}$ and ${url_{\mathrm{part}}}$ respectively. 
For illustration, against 'Netflix', our algorithm obfuscated a new domain \textcolor{purple}{'netflix.hd'}, here the domain is 'hd, whereas 'netflix' is a subdomain. 
Similarly, our method synthesized four ${url_{\mathrm{part}}}$ i.e., \textcolor{purple}{`hd-netflix’,} \textcolor{purple}{‘netflix-hd’},\textcolor{purple}{‘hdnetflix’} and \textcolor{purple}{‘netflixhd’} against ‘netflix’.
To generate (${url_{\mathrm{swap}}}$), the different word tokens of the domain are swapped to obtain a new domain name. For example, for domain 'bankofamerica', our algorithm produces five new domain names, i.e., \textcolor{purple}{bankamericaof}, \textcolor{purple}{ofbankamerica}, \textcolor{purple}{americabankof}, \textcolor{purple}{americaofbank}, \textcolor{purple}{ofamericabank}. 
Lastly, for repetition our algorithm repeated $d$, $n_t$ times to generate a new URL, e.g.,  \textcolor{purple}{‘netflix-netflix’} when $n_t$=2.  
\par {\textbf{(iii) Path Adversary.}}
\label{pathadversary}
Path adversary obfuscated the directory pointing to the web resource of the URL $x$ as manifested by Algorithm~\ref{path}. 
This adversary took the $url_t$, path ($p$), $d$, $x$, ${W_{\mathrm{emb}}}$, list of path extension ${lst_{\mathrm{exe}}}$ and ${url_{\mathrm{part}}}$ as $d_n$ from domain adversary as input and returned the list of path adversaries $P_u$. 
The adversary iteratively substituted each token $loc$ in the path $p$ of $x$ with $d$ to create $p_{\mathrm{dm}}$, e.g., for $x=$ ‘hxxps://www.credit-suisse.com/au/en/\textcolor{teal}{private-banking/contact-us.html}’ is  transformed to ‘hxxps://www.credit-suisse.com/au/en/ \textcolor{purple}{credit-suisse}/private-banking/contact-us.html’.  
Additionally, to create ${path_{\mathrm{exe}}}$, our method swapped the original extension $exe$ of the path with a malicious extension, e.g., the above path changes to `/private-banking/contact-us\textcolor{purple}{.bin}'. 
Finally to generate path adversary, we replaced the original domain $d$ with  $i \in d_n$ and the original path $p$ with $p_{\mathrm{dm}}$ and $p_{\mathrm{exe}}$ successively.
\par{\textbf{(iii) TLD Adversary.}}
TLD adversary changed the original TLD $t$ of $x$ with malicious TLD. For example, 'https://www.google.\textcolor{teal}{com}' was changed to 'https://www.google.\textcolor{purple}{icu}'.
\par{\textbf{(iv) Post Process.}} It reconstructed a URL $x$ from its updated tokens $url_t$ using URL unparse \cite{urllib:online} and produced adversarial candidates $Adv_\mathrm{cand}$. Then it analysed $Adv_\mathrm{cand}$ for their validity using the RFC3986 standard. Consequently, it discarded all the URLs in $Adv_\mathrm{cand}$ that did not follow the generic syntax of a URL provided by the RFC 3986 standard. For example, \url{hxxps://www.2netflix.com} was generated as an adversarial candidate. However, RFC 3986 does not allow the domain names starting from a number; hence this adversary was discarded.
In this way, only valid URLs were selected as Adversarial URLs (AU).
\subsubsection{Experimental Setup}
We used the following experimental setup to generate adversarial examples: 

\textbf{(i) Seed Dataset.} 
\label{seed}
For our experiments, we have considered a seed dataset `S' containing a set of $27467$ legitimate URLs obtained by crawling the top 100 most frequent global phishing target websites \cite{OpenPhish80:online} such as Netflix, eBay, PayPal, Facebook, Amazon. 
Moreover, we only considered the URLs that pointed to the web pages of these websites that contained a $<$form$>$ tag. We did it because we wanted to generate adversarial URLs against the webpages that intend to obtain private information from the users such as username and credentials \cite{Chiew2018}. For example, our list of URLs did not contain `https://www.paypal.com/au/business' because it did not ask for any information from the user. However, we generated adversarial URLs against  \url{`https://www.paypal.com/bizsignup/#/checkAccount'} as it prompts users to interact with the webpage.
The complete list of these URLs is published online \cite{RobustEVALMLPU}.

\textbf{(ii) Word Embedding and other lists.}
We used "English-subword" pre-trained embedding from FastText \cite{mikolov2018advances} as an input.
The rationale behind using this embedding was two-tier. Firstly, this embedding was trained over a Common Crawl database \cite{CommonCr58:online}. 
This database contained gigantic amounts of data obtained by crawling the web for the last seven years. Hence, we believe that word vectors trained on this dataset included contextual relationships between URL and webpage content. 
Secondly, a URL is a sequence of characters and domain names might not be an English word but merely a random sequence of characters. Therefore, the sub-word embedding was suitable for handling Out Of Vocabulary (OOV) words. 
${lst_{\mathrm{tld}}}$ was obtained from \cite{TheTop2022:online} while ${lst_{\mathrm{exe}}}$ was gather from \cite{Filetype32:online}. These lists provides a comprehensive and updated list of TLD and malicious executable respectively.
We fixed semantically related words $n$ to 20 after conducting a pilot study (using $n$ from 10 to 50 with an increment of 5) and analysing the semantic similarity score between the original domain and a related word. We observed that choosing $n >20$ reduces the semantic similarity score to $<60\%$. Lastly, we used validators library \cite{validato72:online} that provides python implementation of RFC 3986 \cite{RFC3986U96:online} standard to select valid $Adv_\mathrm{cand}$. 
\subsection{Reproducing Baseline ML Models}
We were unable to find publicly available codes and datasets for traditional MLPU models proposed by the state-of-the-art. 
Therefore, we reproduced traditional MLPU models by using the most frequent features and classifiers used in the state-of-the-art  (See Table \ref{stateofart} for more details).
For deep learning methods, 
we used the reproducible code of URLNET \cite{urlnet}, EXPOSE \cite{Saxe2017Expose} and LSTM \cite{bahnsen2017classifying} published on GitHub repository \cite{Antimalw45:online}, \cite{joshsaxe70:online} and \cite{GitHubch61:online} respectively. 
\subsubsection{Experimental Setup}
All the experiments ran on a computing cluster with 32 CPU cores with 128GB of RAM and Tesla V100 GPU.
For training the ML models, we used the following setup.
\begin{table}[!tb]
\caption{Datasets used for training and validating ML models}
\label{dataset}
\begin{tabular}{c|c|c}
\hline
\centering
\multirow[c]{5}{8em}{\centering\textbf{Data Sources}}&\textbf{Legitimate} & \textbf{Phishing }\\ 
\cline{2-3}
& Dmoz\cite{URLClass44:online} &PhishTank\cite{PhishTan46:online}\\
 &ISCXURL2016\cite{URL2016D95:online}  &ISCXURL2016\cite{URL2016D95:online} \\
 &Alexa \cite{AlexaTop45:online} web-crawl & OpenPhish \cite{OpenPhish80:online}\\
 &Phish-storm \cite{PhishSto33:online}&Phish-storm \cite{PhishSto33:online}\\
\hline
\textbf{Dataset size}& \textbf{$96693$} & \textbf{$96693$}\\
\hline
\end{tabular}
\end{table}
{\textbf(i) {URL dataset.}}
A large URL dataset of $193,386$ URLs was collected using multiple sources mentioned in Table~\ref{dataset}. We have also published this dataset \cite{RobustEVALMLPU} for research reproducibility. 
Instead of directly using Alexa dataset \cite{AlexaTop45:online}, we crawled 1500 unique domains from this dataset to obtain relative URLs because the Alexa dataset only provides the list of index pages of the websites which creates a bias in the data as highlighted by Shirazi, H., Bezawada, B., and Ray \cite{Shirazi2018KnowThyDomainName}.
For example, `www.netflix.com' is in the Alexa dataset, while the relative URL `www.netflix.com/au/login' is not present. 
On the other hand, phishing feed providers such as Phishtank \cite{PhishTan46:online}, or OpenPhish \cite{OpenPhish80:online} provide lists of complete phishing URLs. 
For instance, \url{hxxps://handaummail.000webhostapp.com/shu/DHL-NEW/D2017HL/u.php} is a phishing URL in OpenPhish \cite{OpenPhish80:online} which contains the full path of phishing resource in the URL. 
After collecting the datasets from the data sources, we removed duplicates and constructed a balanced dataset \inserted{(by considering legitimate URLs from our crawled dataset equal to the size of phishing URL dataset i.e., 96693)} to avoid biases in results (such as learning one class more accurately than the other or overfitting).
\begin{table*}[!tb]
  \caption{Details of traditional ML classifier features ($min-df$ is a threshold for removing infrequent word in the corpus, $n$ represents n-gram range)}
 \label{features}
\begin{small}
\centering
\resizebox{\textwidth}{!}
 {\begin{tabular}{>{\raggedright\arraybackslash}m{20mm}|m{40mm}|m{40mm}|m{5mm}|m{20mm}}
\hline
\multicolumn{1}{>{\centering\arraybackslash}m{15mm}|}{\textbf{Feature type}} 
&\multicolumn{1}{>{\centering\arraybackslash}m{20mm}|}{\textbf{Description}} 
    & \multicolumn{1}{>{\centering\arraybackslash}m{40mm}|}{\textbf{Features (F)}} 
    & \multicolumn{1}{>{\centering\arraybackslash}m{6mm}|}{\textbf{No of F}}
	 & \multicolumn{1}{>{\centering\arraybackslash}m{10mm}}{\textbf{Study}}\\
\hline
\textbf{Basic Lexical}& Represent the statistical properties of a URL string.&
Count of special characters in URL and each part of URL, TLD in arguments, No of parameters, File extension, No of different characters, Ratio of no of digits in domain name to its length, Ratio of number of consonants to URL length, Ratio of number of consonants to number of vowels, Shortening service, URL entropy&134&\cite{crackingclassifier,Mamun2016LexicalMaliciousURL,Le2011PhishDef,Jain2018phishingwebsitedetection}
\\
\hline
\textbf{Bag of Words (BoW)}& Represent the most frequent words in a URL dataset&min-df=0.0001, n=(1,1)&6458&\cite{kolari2006svms,patgiri2019empirical}  \\
\hline
\textbf{BoW of URL parts}& Represent the BoW in each part of the URL.& min-df=0.0001, n=(1,1)&10237&\cite{ma2009identifying,Vanhoenshoven2016MaliciousURLs,ma2011LearningtoDetectMaliciousURLs}. 
 \\
\hline
\textbf{Bigram}& Represent the closest context of a word and constitute two frequent consecutive words in a URL dataset.& min-df=0.0001,n=(1,2) &16222&\cite{Blum2010Lexical, Feroz2014Examination} \\
\hline
\textbf{Char n-grams}& Represent the most common n character sequences in a URL dataset.&min-df=0.0001 , n=(3-8)&60929&\cite{verma2017s,Darling2015lexicalmaliciousURL}\\
\hline
\textbf{External}& Represent the quality attributes of a URL computed by an external third-party.&
Domain has valid Sender Policy Framework (SPF),
Domain presence in RBL (Real-time Blackhole List),
Response Time, 
ASN number associated with the IP,
Return the country associated with IP,
Return DNS Pointer Record (PTR) associated with IP,
Domain activation time, Domain Expiration time,
Count of resolved IPs, 
Count of resolved Name servers,
Number of Resolved Mail (MX) servers,
Time-to-live (TTL) value associated with hostname,
Valid secure Sockets Layer (SSL) certificate,
Google indexed URL, 
Google indexed domain&15&\cite{Le2011PhishDef}\\
\hline
\end{tabular}}
\end{small}
\end{table*}

{\textbf{(ii) Feature Extraction.}}
For training traditional ML classifiers, popular features used by the state-the-art MLPU models were extracted (see Table~\ref{features} for details).
External features extraction took an average of 16 seconds per URL; the latency was due to external queries \cite{Le2011PhishDef}. 
We extracted these features using multiprocessing code with 32 CPUs to deal with this latency effectively.
For deep learning methods, URLNET \cite{urlnet} transformed URLs into two types of representations: word and character vectors while EXPOSE \cite{Saxe2017Expose} only used character vectors. 
The word and character level vectors were input to the deep neural network that extracts proper pattern and classification weights.

\textbf{(iii) Hyperparameter and training Setting.}
\label{hypertuning}
To select optimal traditional ML models, we applied Bayesian optimization \cite{snoek2012practical} using hyperopt library \cite{bergstra2013making}.
We chose bayesian optimisation because it is robust to noisy objective function evaluations \cite{wang2013bayesian}.
We utilised the average Matthew Correction Coefficient (MCC) of 10-fold cross-validation with stratified sampling \cite{pmlr-v48-liberty16} (that retains a balanced class distribution in each fold) and early stopping criteria to select the optimal parameters.
MCC was used to select the optimal model since MCC explicitly considers all classes \cite{luque2019impact} and is resilient against the
imbalance datasets \cite{luque2019impact,chicco2020advantages,bergstra2013making}. 
MCC value ranges from -1 to 1, a value close to -1, 0, 1 depicts a poor model (misclassify both classes), random model (classify both classes randomly) and a good model (classify both classes well), respectively.
In contrast, for deep learning, we used the hyper-parameters provided by the original paper or the GitHub repositories \cite{Saxe2017Expose,urlnet, bahnsen2017classifying,Haplophysh}.

\textbf{(iv) Original Performance Measures.}
\label{origperf}
To report the actual performance of the MLPU systems, we have used MCC, accuracy (ACC), AUC, FPR and FNR.
Previous studies \cite{luque2019impact,chicco2020advantages,bergstra2013making} have shown that these metrics summarise the overall performance of ML models better than other measures especially .

\subsection{Evaluate URLBUG Realizability (RQ1)}
\label{realizabilityevaluation}
To answer RQ1, we firstly studied the computational effort required to create AUs for a single target URL. Then we analysed the annual cost of registering them.
Suppose the computational effort and annual registration cost are low. In that case, less effort is required to generate AUs, and they pose a real threat to the MLPU models or vice-versa. Lastly, to ensure the generated URLs were adversarial (deceptive) in nature, we validated the generated URLs against the definition of deceptive URLs formulated by the previous studies \cite{Le2011PhishDef, IdentifyConfusion2020CHI}.
According to the definition, the generated URL should obfuscate the identity of the victim domain by either containing its domain as a part of the URL or should a be typo-squatting version (look similar but spelt differently) of a target domain.
\subsubsection{Experimental Setup}
To answer RQ1, we used the following experimental setup:
\textbf{(i) Evaluate Time (RQ1.1).}
Our AUs were generated in a Core i7 CPU 2.2Hz with a 16GB RAM system. Using the python default time library, we computed the time taken to generate an $AU$ against a URL $x$.
\par
\textbf{(ii) Evaluate Cost (RQ1.2).}
We queried the $AU$ availability for registration and annual prices using GoDaddy \cite{httpsdev96:online}. GoDaddy is one of the most extensive domain names registration services.
\begin{table*}[!tb]
  \caption{Evaluation Metrics for Deceptiveness of a URL (the valid values for each metric can be [True i.e., deceptive or False i.e., legitimate])}
  \label{RQ1matrix}
\begin{small}
\centering
\begin{tabular}{>{\raggedright\arraybackslash}m{8mm}|m{30mm}|m{50mm}|m{80mm}}
\hline
\multicolumn{1}{>{\centering\arraybackslash}m{8mm}|}{\textbf{Type}} 
&\multicolumn{1}{>{\centering\arraybackslash}m{15mm}|}{\textbf{Metric}} 
    & \multicolumn{1}{>{\centering\arraybackslash}m{40mm}|}{\textbf{Description and Rationale}} 
    & \multicolumn{1}{>{\centering\arraybackslash}m{55mm}}{\textbf{Experimental Setup}}\\
\hline
\multirow{2}{*}{\rotatebox{90}{\textbf{Legally }}}\multirow{2}{*}{\rotatebox{90}{\textbf{ disputed}}}&\textbf{Blacklisted}& This metric checks whether the generated URL is already blacklisted as phishing URL in the past \cite{oest2020phishtime}.
&
We used four blacklists: Google Safe Browsing \cite{SafeBrow76:online}, VirusTotal \cite{virustotal}, PhishTank \cite{PhishTan46:online} and Openphish \cite{OpenPhish80:online}. These are well-known blacklists that contain information of a URL previously detected as phishing.\\
\cline{2-4}
&\textbf{Prohibited} & This metric assist to discover that whether the URL is subjected to legal disputes. &We queried WHOIS \cite{Whoiscom89:online} server to obtain information about the status \cite{EPPStatu9} of the URL.\\
 \hline
 \multirow{3}{*}{\rotatebox{90}{\textbf{Intentionally}}}
 \multirow{3}{*}{\rotatebox{90}{\textbf{Protected}}}&\textbf{Privacy Protected} &This metric represent known threat to Brands and  determines whether the hostname is registered by a privacy protected companies such as MarkMonitor or Safe names Ltd \cite{gorman2015reclaiming,yue2010bogusbiter} preserve the identity of a brand.
 & To determine whether the hostname is registered by a privacy protected company , we studied the registrar of each hostname by querying WHOIS database.\\
\cline{2-4}
&\textbf{Registered by seed} & 
This metric points out a known deceptive variant of the seed URL registered by the original brand name to protect its integrity.
&To obtain this information, we examined the registrar information of the hostnames using the WHOIS service.\\
\hline
\multirow{8}{*}{\rotatebox{90}{\textbf{Redirected}}}
&\textbf{Auctioned} & This metric helps to find out whether the URL belongs to a genuine website or redirects to a webpage that auctions 
its hostname. For example, \url{hxxps://aboBank.com} redirected to \url{hxxps://www. hugedomains.com/} that auctions this domain for \$5090. & To check for hostname being auctioned, we examined the landing page text for keywords such as sale, auction, purchase or buy.\\
\cline{2-4}
&\textbf{Redirected to seed} &This metrics indicate whether the generated 
URL redirects request to the original seed URL.  & For getting this information, we checked the redirect history and the final URL destination of the URL using urllib library. Then we compared the final URL hostname with the original seed hostname.\\
\cline{2-4}
&\textbf{Redirected to Blacklisted} & These metrics ascertain that the generated URL redirects to a blocked URL. & We scanned the redirected URLs against four blacklists (Google Safe Browsing \cite{SafeBrow76:online}, VirusTotal \cite{virustotal}, PhishTank \cite{PhishTan46:online} and Openphish \cite{OpenPhish80:online}).\\
\cline{2-4}
&\textbf{Redirected to Unknown} &This metric examines whether the generated URL redirects to an unknown destination (exclusive of blacklisted or seed hostnames).& We consider the destination of the generated URL unknown if it did not redirect to either seed URL or blacklist URL.\\
\hline
\end{tabular}
\end{small}
\end{table*}
\par\textbf{(iii) Evaluate Deception (RQ1.3).}
To ensure the $AU$ were adversarial \textit{deceptive} in nature, \inserted{our perturbation were constrained to either the target domain name should be a part of the adversarial URL or edit distance (number of character transformations) between real and adversarial URL must be minimal \cite{Gao2018DeepWordBug} (less than 10\% of the total length of the real URL). These constraints are essential for satisfying the semantics of deceptive URL \cite{Le2011PhishDef}. For ensuring this we used the string matching approach to find the target domain is present in the generated deceptive URL and for typo-squatting constraint, we have used Levenshtein distance equal to 1 \cite{textdist70:online}}. 
We have selected these approaches for their popularity in the related text-classification field for generating adversaries \cite{morris2020textattack}.

Further, we assessed the deceptiveness of generated URLs that were already registered based on the metrics shown in Table~\ref{RQ1matrix}.
We designed this matrix after studying several studies \cite{oest2020phishtime,IdentifyConfusion2020CHI,albakry2020url,Le2011PhishDef,Mamun2016LexicalMaliciousURL} that examined the deceptiveness of the URL from users and ML perspective.
The metrics evaluated the deceptiveness of a URL based on three types of measures.
Firstly, it checked whether the URL was legally disputed, i.e., it was blocked in the past, or its WHOIS status code indicates it as a prohibited URL.
WHOIS is the largest domain database that contains information about domain registrars and their activation status.
We have considered the activation status prohibited if the WHOIS query returns any of the following statuses: serverrenewprohibited, serverTransferProhibited, serverUpdateProhibited, clientRenewProhibited. These status codes are uncommon and depict a legal dispute against the hostname \cite{EPPStatu9}.
We selected this measure as an indication of deception because previous studies \cite{oest2020phishtime,mcgrath2008behind} have suggested that these measures were adequate to detect deceptive URLs.
Intentionally protected represented those variants of the seed URL that were either registered by identity theft protection service such as MarkMonitor \cite{MarkMoni78} or by the original owner of the domain to protect their identity. Such variants were considered deceptive because they represented a known threat to the brand's identity and, if they were not registered, could be used for malicious purposes.
Lastly, we considered redirection as the prior studies \cite{oest2020sunrise, Le2011PhishDef} have identified that redirection had been used as one of the critical weapons by phishing attacks. 
This metrics was computing using the experimental setup given by third column of Table~\ref{RQ1matrix}.

\subsection{Evaluate Adversarial Robustness}
To assess the robustness of the pre-trained models, we subjected them to our Adversarial URL dataset $Adv_\mathrm{data}$ containing 27467 benign seed URLs and 1,515,750 adversarial URLs. We first generated AE dataset using the Seed URL Dataset S and then used $30,000$ adversarial URLs (that were not already registered) to test the trained MLPU models.
\begin{figure}[!tb]
\centering{\includegraphics[width=0.8\columnwidth]{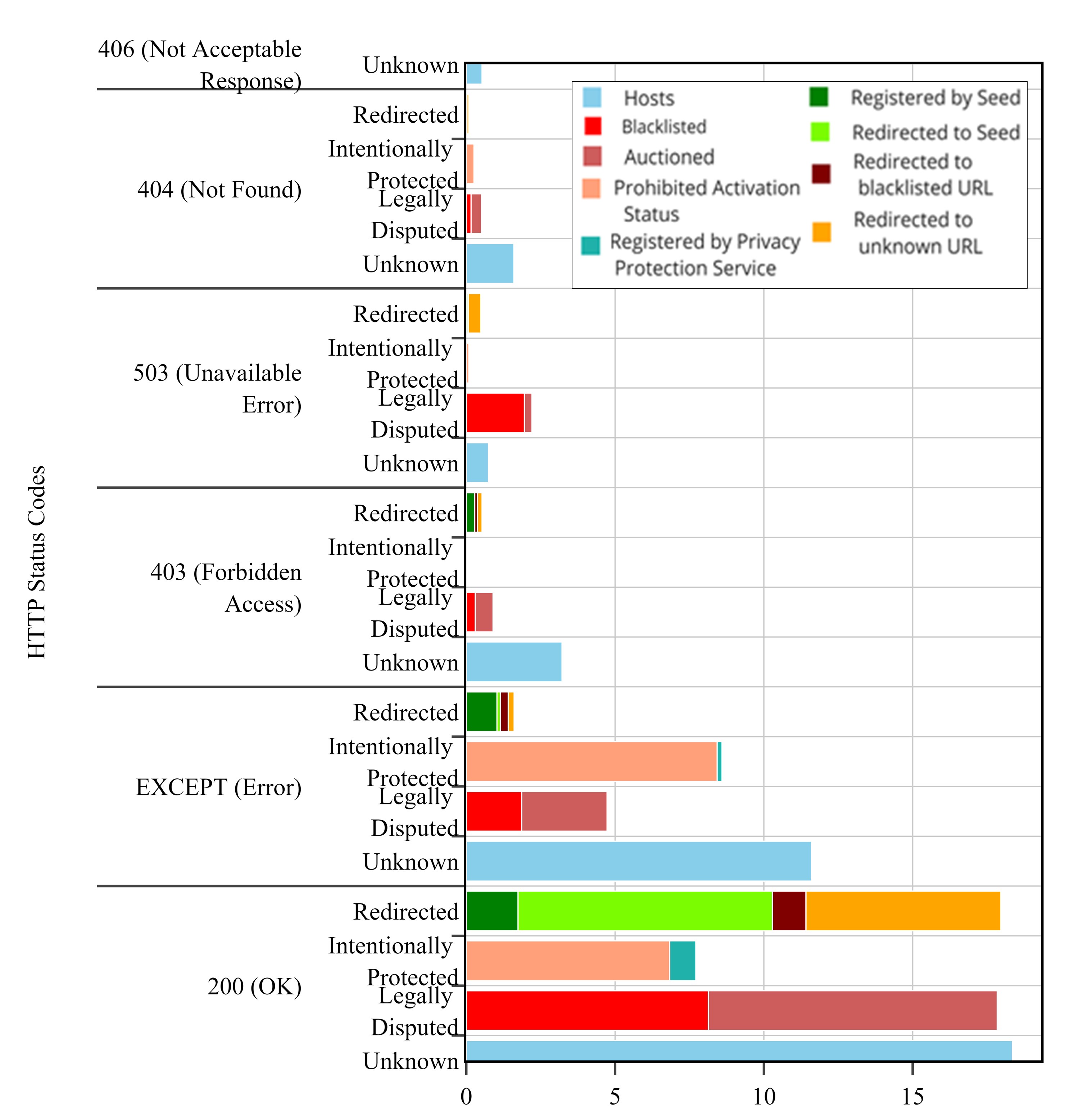}}
\centering\caption{Evaluation of deceptiveness of generated URLs}
\label{AnalysisRQ1}
\end{figure}
\begin{table*}[tb]
\caption{Original Performances of the MLPU models versus Adversarial Performance [Red, green, blue and orange color highlights the models with ($<$0.5), ($>$0.90), $<$0.90 and $>$0.50 and $0$ performance metric value respectively]}
\label{RQ1Table}
 \begin{small}
\begin{tabular}{m{30mm}|m{15mm}|m{7.5mm}|m{7.5mm}|m{7.5mm}|m{7.5mm}|m{7.5mm}|m{7.5mm}|m{7.5mm}|m{7.5mm}|m{7.5mm}|m{7.5mm}}
    \hline
\multicolumn{2}{c|}{\textbf{Models}}&\multicolumn{5}{c|}{\textbf{Original Performance}}&\multicolumn{5}{c}{\textbf{Adversarial Performance}}\\
\hline
\textbf{Features}&\textbf{Classifier}&\textbf{MCC}&\textbf{ACC}&\textbf{AUC}&\textbf{FPR}&\textbf{FNR}&\textbf{MCC}&\textbf{ACC}&\textbf{AUC}&\textbf{FPR}&\textbf{FNR}\\
\hline
\multirow{7}{*}{\textbf{Basic Lexical}}&RF&{\cellcolor{white!25}{0.910}}&{\cellcolor{white!25}{95.99\%}}&{\cellcolor{white!25}{99.10\%}}&{\cellcolor{white!25}{2.34\%}}&{\cellcolor{white!25}{4.81\%}}&{\cellcolor{white!25}{-0.035}}&{\cellcolor{white!25}{15.09\%}}&{\cellcolor{white!25}{42.98\%}}&{\cellcolor{white!25}{28.67\%}}&{\cellcolor{white!25}{85.36\%}}\\
&DT&{\cellcolor{white!25}{0.844}}&{\cellcolor{white!25}{92.96\%}}&{\cellcolor{white!25}{92.47\%}}&{\cellcolor{white!25}{6.00\%}}&{\cellcolor{white!25}{7.53\%}}&{\cellcolor{white!25}{0.034}}&{\cellcolor{white!25}{51.66\%}}&{\cellcolor{white!25}{59.46\%}}&{\cellcolor{white!25}{32.62\%}}&{\cellcolor{white!25}{48.47\%}}\\
&SVM&{\cellcolor{white!25}{0.747}}&{\cellcolor{white!25}{88.85\%}}&{\cellcolor{white!25}{94.52\%}}&{\cellcolor{white!25}{5.28\%}}&{\cellcolor{white!25}{13.93\%}}&{\cellcolor{white!25}{0.000}}&{\cellcolor{white!25}{99.21\%}}&{\cellcolor{white!25}{50.00\%}}&{\cellcolor{white!25}{100.00\%}}&{\cellcolor{white!25}{0.00\%}}\\
&LR&{\cellcolor{white!25}{0.745}}&{\cellcolor{white!25}{88.40\%}}&{\cellcolor{white!25}{94.76\%}}&{\cellcolor{white!25}{10.09\%}}&{\cellcolor{white!25}{12.31\%}}&{\cellcolor{white!25}{0.019}}&{\cellcolor{white!25}{51.21\%}}&{\cellcolor{white!25}{55.32\%}}&{\cellcolor{white!25}{40.50\%}}&{\cellcolor{white!25}{48.86\%}}\\
&KNN&{\cellcolor{white!25}{0.879}}&{\cellcolor{white!25}{94.58\%}}&{\cellcolor{white!25}{93.74\%}}&{\cellcolor{white!25}{3.65\%}}&{\cellcolor{white!25}{6.26\%}}&{\cellcolor{white!25}{-0.031}}&{\cellcolor{white!25}{14.17\%}}&{\cellcolor{white!25}{43.94\%}}&{\cellcolor{white!25}{25.81\%}}&{\cellcolor{white!25}{86.31\%}}\\
&XGB&{\cellcolor{white!25}{0.923}}&{\cellcolor{white!25}{96.57\%}}&{\cellcolor{white!25}{99.30\%}}&{\cellcolor{white!25}{2.28\%}}&{\cellcolor{white!25}{3.97\%}}&{\cellcolor{white!25}{0.001}}&{\cellcolor{white!25}{28.81\%}}&{\cellcolor{white!25}{50.25\%}}&{\cellcolor{white!25}{27.96\%}}&{\cellcolor{white!25}{71.54\%}}\\
&LGBM&{\cellcolor{white!25}{0.922}}&{\cellcolor{white!25}{96.52\%}}&{\cellcolor{white!25}{99.25\%}}&{\cellcolor{white!25}{2.39\%}}&{\cellcolor{white!25}{4.00\%}}&{\cellcolor{white!25}{-0.002}}&{\cellcolor{white!25}{27.70\%}}&{\cellcolor{white!25}{49.52\%}}&{\cellcolor{white!25}{28.32\%}}&{\cellcolor{white!25}{72.65\%}}\\
\hline
\multirow{7}{*}{\textbf{BoW}}
&RF&{\cellcolor{white!25}{0.935}}&{\cellcolor{white!25}{96.75\%}}&{\cellcolor{white!25}{99.36\%}}&{\cellcolor{white!25}{2.39\%}}&{\cellcolor{white!25}{3.25\%}}&{\cellcolor{white!25}{0.020}}&{\cellcolor{white!25}{58.24\%}}&{\cellcolor{white!25}{55.47\%}}&{\cellcolor{white!25}{47.35\%}}&{\cellcolor{white!25}{41.72\%}}\\
&DT&{\cellcolor{white!25}{0.917}}&{\cellcolor{white!25}{95.85\%}}&{\cellcolor{white!25}{95.98\%}}&{\cellcolor{white!25}{3.84\%}}&{\cellcolor{white!25}{4.15\%}}&{\cellcolor{white!25}{0.006}}&{\cellcolor{white!25}{52.69\%}}&{\cellcolor{white!25}{51.62\%}}&{\cellcolor{white!25}{49.47\%}}&{\cellcolor{white!25}{47.29\%}}\\
&SVM&{\cellcolor{white!25}{0.915}}&{\cellcolor{white!25}{95.75\%}}&{\cellcolor{white!25}{99.02\%}}&{\cellcolor{white!25}{2.77\%}}&{\cellcolor{white!25}{4.25\%}}&{\cellcolor{white!25}{0.013}}&{\cellcolor{white!25}{59.85\%}}&{\cellcolor{white!25}{53.48\%}}&{\cellcolor{white!25}{53.00\%}}&{\cellcolor{white!25}{40.05\%}}\\
&LR&{\cellcolor{white!25}{0.916}}&{\cellcolor{white!25}{95.78\%}}&{\cellcolor{white!25}{99.04\%}}&{\cellcolor{white!25}{2.80\%}}&{\cellcolor{white!25}{4.22\%}}&{\cellcolor{white!25}{0.011}}&{\cellcolor{white!25}{59.99\%}}&{\cellcolor{white!25}{53.02\%}}&{\cellcolor{white!25}{54.06\%}}&{\cellcolor{white!25}{39.90\%}}\\
&KNN&{\cellcolor{white!25}{0.914}}&{\cellcolor{white!25}{95.60\%}}&{\cellcolor{white!25}{95.60\%}}&{\cellcolor{white!25}{1.11\%}}&{\cellcolor{white!25}{4.40\%}}&{\cellcolor{white!25}{-0.014}}&{\cellcolor{white!25}{25.32\%}}&{\cellcolor{white!25}{46.58\%}}&{\cellcolor{white!25}{31.80\%}}&{\cellcolor{white!25}{75.03\%}}\\
&XGB&{\cellcolor{white!25}{0.941}}&{\cellcolor{white!25}{97.03\%}}&{\cellcolor{white!25}{99.47\%}}&{\cellcolor{white!25}{2.46\%}}&{\cellcolor{white!25}{2.97\%}}&{\cellcolor{white!25}{0.016}}&{\cellcolor{white!25}{59.52\%}}&{\cellcolor{white!25}{54.36\%}}&{\cellcolor{white!25}{50.88\%}}&{\cellcolor{white!25}{40.40\%}}\\
&LGBM&{\cellcolor{white!25}{0.945}}&{\cellcolor{white!25}{97.23\%}}&{\cellcolor{white!25}{99.49\%}}&{\cellcolor{white!25}{2.19\%}}&{\cellcolor{white!25}{2.77\%}}&{\cellcolor{white!25}{0.011}}&{\cellcolor{white!25}{59.52\%}}&{\cellcolor{white!25}{53.13\%}}&{\cellcolor{white!25}{53.36\%}}&{\cellcolor{white!25}{40.38\%}}\\
\hline
\multirow{7}{*}{\textbf{Bigram}}&RF&{\cellcolor{white!25}{0.937}}&{\cellcolor{white!25}{96.85\%}}&{\cellcolor{white!25}{99.39\%}}&{\cellcolor{white!25}{2.29\%}}&{\cellcolor{white!25}{3.15\%}}&{\cellcolor{white!25}{0.023}}&{\cellcolor{white!25}{57.35\%}}&{\cellcolor{white!25}{56.42\%}}&{\cellcolor{white!25}{44.52\%}}&{\cellcolor{white!25}{42.64\%}}\\
&DT&{\cellcolor{white!25}{0.918}}&{\cellcolor{white!25}{95.91\%}}&{\cellcolor{white!25}{96.06\%}}&{\cellcolor{white!25}{3.86\%}}&{\cellcolor{white!25}{4.09\%}}&{\cellcolor{white!25}{0.039}}&{\cellcolor{white!25}{54.19\%}}&{\cellcolor{white!25}{60.96\%}}&{\cellcolor{white!25}{32.16\%}}&{\cellcolor{white!25}{45.92\%}}\\
&SVM&{\cellcolor{white!25}{0.924}}&{\cellcolor{white!25}{96.16\%}}&{\cellcolor{white!25}{99.16\%}}&{\cellcolor{white!25}{2.60\%}}&{\cellcolor{white!25}{3.84\%}}&{\cellcolor{white!25}{0.011}}&{\cellcolor{white!25}{60.18\%}}&{\cellcolor{white!25}{52.94\%}}&{\cellcolor{white!25}{54.42\%}}&{\cellcolor{white!25}{39.70\%}}\\
&LR&{\cellcolor{white!25}{0.923}}&{\cellcolor{white!25}{96.15\%}}&{\cellcolor{white!25}{99.18\%}}&{\cellcolor{white!25}{2.66\%}}&{\cellcolor{white!25}{3.85\%}}&{\cellcolor{white!25}{0.013}}&{\cellcolor{white!25}{60.11\%}}&{\cellcolor{white!25}{53.43\%}}&{\cellcolor{white!25}{53.36\%}}&{\cellcolor{white!25}{39.78\%}}\\
&KNN&{\cellcolor{white!25}{0.919}}&{\cellcolor{white!25}{95.88\%}}&{\cellcolor{white!25}{95.88\%}}&{\cellcolor{white!25}{1.27\%}}&{\cellcolor{white!25}{4.12\%}}&{\cellcolor{white!25}{0.031}}&{\cellcolor{white!25}{48.03\%}}&{\cellcolor{white!25}{58.73\%}}&{\cellcolor{white!25}{30.39\%}}&{\cellcolor{white!25}{52.14\%}}\\
&XGB&{\cellcolor{white!25}{0.944}}&{\cellcolor{white!25}{97.20\%}}&{\cellcolor{white!25}{99.51\%}}&{\cellcolor{white!25}{2.37\%}}&{\cellcolor{white!25}{2.80\%}}&{\cellcolor{white!25}{0.022}}&{\cellcolor{white!25}{59.91\%}}&{\cellcolor{white!25}{56.14\%}}&{\cellcolor{white!25}{47.70\%}}&{\cellcolor{white!25}{40.03\%}}\\
&LGBM&{\cellcolor{white!25}{0.936}}&{\cellcolor{white!25}{96.78\%}}&{\cellcolor{white!25}{99.41\%}}&{\cellcolor{white!25}{2.44\%}}&{\cellcolor{white!25}{3.22\%}}&{\cellcolor{white!25}{0.016}}&{\cellcolor{white!25}{61.13\%}}&{\cellcolor{white!25}{54.30\%}}&{\cellcolor{white!25}{52.65\%}}&{\cellcolor{white!25}{38.76\%}}\\
\hline
\multirow{7}{*}{\textbf{Char n-gram}}&RF&{\cellcolor{white!25}{0.956}}&{\cellcolor{white!25}{97.80\%}}&{\cellcolor{white!25}{99.71\%}}&{\cellcolor{white!25}{2.05\%}}&{\cellcolor{white!25}{2.20\%}}&{\cellcolor{white!25}{0.020}}&{\cellcolor{white!25}{57.53\%}}&{\cellcolor{white!25}{55.46\%}}&{\cellcolor{white!25}{46.64\%}}&{\cellcolor{white!25}{42.44\%}}\\
&DT&{\cellcolor{white!25}{0.923}}&{\cellcolor{white!25}{96.17\%}}&{\cellcolor{white!25}{94.67\%}}&{\cellcolor{white!25}{3.86\%}}&{\cellcolor{white!25}{3.83\%}}&{\cellcolor{white!25}{0.026}}&{\cellcolor{white!25}{58.47\%}}&{\cellcolor{white!25}{57.16\%}}&{\cellcolor{white!25}{44.17\%}}&{\cellcolor{white!25}{41.51\%}}\\
&SVM&{\cellcolor{white!25}{0.952}}&{\cellcolor{white!25}{97.58\%}}&{\cellcolor{white!25}{99.62\%}}&{\cellcolor{white!25}{2.74\%}}&{\cellcolor{white!25}{2.42\%}}&{\cellcolor{white!25}{0.030}}&{\cellcolor{white!25}{61.68\%}}&{\cellcolor{white!25}{58.25\%}}&{\cellcolor{white!25}{45.23\%}}&{\cellcolor{white!25}{38.27\%}}\\
&LR&{\cellcolor{white!25}{0.960}}&{\cellcolor{white!25}{97.98\%}}&{\cellcolor{white!25}{99.75\%}}&{\cellcolor{white!25}{1.94\%}}&{\cellcolor{white!25}{2.02\%}}&{\cellcolor{white!25}{0.032}}&{\cellcolor{white!25}{61.41\%}}&{\cellcolor{white!25}{58.82\%}}&{\cellcolor{white!25}{43.82\%}}&{\cellcolor{white!25}{38.55\%}}\\
&KNN&{\cellcolor{white!25}{0.923}}&{\cellcolor{white!25}{96.09\%}}&{\cellcolor{white!25}{97.99\%}}&{\cellcolor{white!25}{1.78\%}}&{\cellcolor{white!25}{3.91\%}}&{\cellcolor{white!25}{0.050}}&{\cellcolor{white!25}{52.32\%}}&{\cellcolor{white!25}{63.88\%}}&{\cellcolor{white!25}{24.38\%}}&{\cellcolor{white!25}{47.86\%}}\\
&\textbf{XGB}&{\cellcolor{white!25}{{0.972}}}&{\cellcolor{white!25}{98.59\%}}&{\cellcolor{white!25}{99.86\%}}&{\cellcolor{white!25}{1.34\%}}&{\cellcolor{white!25}{1.41\%}}&{\cellcolor{white!25}{0.025}}&{\cellcolor{white!25}{60.53\%}}&{\cellcolor{white!25}{56.97\%}}&{\cellcolor{white!25}{46.64\%}}&{\cellcolor{white!25}{39.42\%}}\\
&LGBM&{\cellcolor{white!25}{0.966}}&{\cellcolor{white!25}{98.31\%}}&{\cellcolor{white!25}{99.84\%}}&{\cellcolor{white!25}{1.48\%}}&{\cellcolor{white!25}{1.69\%}}&{\cellcolor{white!25}{0.025}}&{\cellcolor{white!25}{62.53\%}}&{\cellcolor{white!25}{56.76\%}}&{\cellcolor{white!25}{49.12\%}}&{\cellcolor{white!25}{37.37\%}}\\
\hline
\multirow{4}{*}{\textbf{BoW of URL}}
s&RF&{\cellcolor{white!25}{0.893}}&{\cellcolor{white!25}{94.61\%}}&{\cellcolor{white!25}{98.45\%}}&{\cellcolor{white!25}{3.35\%}}&{\cellcolor{white!25}{5.39\%}}&{\cellcolor{white!25}{0.000}}&{\cellcolor{white!25}{12.30\%}}&{\cellcolor{white!25}{0.00\%}}&{\cellcolor{white!25}{0.00\%}}&{\cellcolor{white!25}{87.70\%}}\\
&DT&{\cellcolor{white!25}{0.874}}&{\cellcolor{white!25}{93.68\%}}&{\cellcolor{white!25}{94.63\%}}&{\cellcolor{white!25}{4.81\%}}&{\cellcolor{white!25}{6.32\%}}&{\cellcolor{white!25}{0.000}}&{\cellcolor{white!25}{3.15\%}}&{\cellcolor{white!25}{0.00\%}}&{\cellcolor{white!25}{0.00\%}}&{\cellcolor{white!25}{96.85\%}}\\
&SVM&{\cellcolor{white!25}{0.878}}&{\cellcolor{white!25}{93.85\%}}&{\cellcolor{white!25}{98.08\%}}&{\cellcolor{white!25}{3.59\%}}&{\cellcolor{white!25}{6.15\%}}&{\cellcolor{white!25}{0.000}}&{\cellcolor{white!25}{15.58\%}}&{\cellcolor{white!25}{0.00\%}}&{\cellcolor{white!25}{0.00\%}}&{\cellcolor{white!25}{84.42\%}}\\
&LR&{\cellcolor{white!25}{0.878}}&{\cellcolor{white!25}{93.87\%}}&{\cellcolor{white!25}{98.11\%}}&{\cellcolor{white!25}{3.77\%}}&{\cellcolor{white!25}{6.13\%}}&{\cellcolor{white!25}{0.000}}&{\cellcolor{white!25}{15.38\%}}&{\cellcolor{white!25}{0.00\%}}&{\cellcolor{white!25}{0.00\%}}&{\cellcolor{white!25}{84.62\%}}\\
\textbf{part}&KNN&{\cellcolor{white!25}{0.801}}&{\cellcolor{white!25}{89.38\%}}&{\cellcolor{white!25}{95.21\%}}&{\cellcolor{white!25}{1.62\%}}&{\cellcolor{white!25}{10.62\%}}&{\cellcolor{white!25}{0.000}}&{\cellcolor{white!25}{41.81\%}}&{\cellcolor{white!25}{0.00\%}}&{\cellcolor{white!25}{0.00\%}}&{\cellcolor{white!25}{58.18\%}}\\
&XGB&{\cellcolor{white!25}{0.901}}&{\cellcolor{white!25}{95.04\%}}&{\cellcolor{white!25}{98.66\%}}&{\cellcolor{white!25}{3.65\%}}&{\cellcolor{white!25}{4.96\%}}&{\cellcolor{white!25}{0.000}}&{\cellcolor{white!25}{33.66\%}}&{\cellcolor{white!25}{0.00\%}}&{\cellcolor{white!25}{0.00\%}}&{\cellcolor{white!25}{66.34\%}}\\
&LGBM&{\cellcolor{white!25}{0.895}}&{\cellcolor{white!25}{94.74\%}}&{\cellcolor{white!25}{98.58\%}}&{\cellcolor{white!25}{3.67\%}}&{\cellcolor{white!25}{5.26\%}}&{\cellcolor{white!25}{0.000}}&{\cellcolor{white!25}{13.53\%}}&{\cellcolor{white!25}{0.00\%}}&{\cellcolor{white!25}{0.00\%}}&{\cellcolor{white!25}{86.47\%}}\\

\hline
\multirow{4}{*}{\textbf{Basic Lexical+}} &RF&{\cellcolor{white!25}{0.943}}&{\cellcolor{white!25}{97.15\%}}&{\cellcolor{white!25}{99.63\%}}&{\cellcolor{white!25}{3.12\%}}&{\cellcolor{white!25}{2.85\%}}&{\cellcolor{white!25}{0.083}}&{\cellcolor{white!25}{69.27\%}}&{\cellcolor{white!25}{71.54\%}}&{\cellcolor{white!25}{26.15\%}}&{\cellcolor{white!25}{30.77\%}}\\
&DT&{\cellcolor{white!25}{0.887}}&{\cellcolor{white!25}{94.34\%}}&{\cellcolor{white!25}{94.34\%}}&{\cellcolor{white!25}{5.98\%}}&{\cellcolor{white!25}{5.66\%}}&{\cellcolor{white!25}{0.047}}&{\cellcolor{white!25}{52.69\%}}&{\cellcolor{white!25}{63.19\%}}&{\cellcolor{white!25}{26.15\%}}&{\cellcolor{white!25}{47.48\%}}\\
&SVM&{\cellcolor{white!25}{0.827}}&{\cellcolor{white!25}{91.27\%}}&{\cellcolor{white!25}{97.21\%}}&{\cellcolor{white!25}{7.85\%}}&{\cellcolor{white!25}{8.73\%}}&{\cellcolor{white!25}{\textbf{0.281}}}&{\cellcolor{white!25}{97.23\%}}&{\cellcolor{white!25}{76.17\%}}&{\cellcolor{white!25}{45.23\%}}&{\cellcolor{white!25}{2.43\%}}\\
&LR&{\cellcolor{white!25}{0.832}}&{\cellcolor{white!25}{91.56\%}}&{\cellcolor{white!25}{97.26\%}}&{\cellcolor{white!25}{6.69\%}}&{\cellcolor{white!25}{8.44\%}}&{\cellcolor{white!25}{\textbf{0.302}}}&{\cellcolor{white!25}{97.27\%}}&{\cellcolor{white!25}{78.29\%}}&{\cellcolor{white!25}{40.99\%}}&{\cellcolor{white!25}{2.42\%}}\\
\textbf{External} &KNN&{\cellcolor{white!25}{0.840}}&{\cellcolor{white!25}{91.73\%}}&{\cellcolor{white!25}{98.28\%}}&{\cellcolor{white!25}{2.60\%}}&{\cellcolor{white!25}{8.28\%}}&{\cellcolor{white!25}{\textbf{0.241}}}&{\cellcolor{white!25}{89.08\%}}&{\cellcolor{white!25}{93.27\%}}&{\cellcolor{white!25}{2.47\%}}&{\cellcolor{white!25}{10.99\%}}\\
&XGB&{\cellcolor{white!25}{0.956}}&{\cellcolor{white!25}{97.78\%}}&{\cellcolor{white!25}{99.75\%}}&{\cellcolor{white!25}{2.27\%}}&{\cellcolor{white!25}{2.22\%}}&{\cellcolor{white!25}{0.080}}&{\cellcolor{white!25}{65.37\%}}&{\cellcolor{white!25}{71.50\%}}&{\cellcolor{white!25}{22.26\%}}&{\cellcolor{white!25}{34.74\%}}\\
&LGBM&{\cellcolor{white!25}{0.957}}&{\cellcolor{white!25}{97.83\%}}&{\cellcolor{white!25}{99.76\%}}&{\cellcolor{white!25}{2.25\%}}&{\cellcolor{white!25}{2.17\%}}&{\cellcolor{white!25}{0.089}}&{\cellcolor{white!25}{67.63\%}}&{\cellcolor{white!25}{73.34\%}}&{\cellcolor{white!25}{20.85\%}}&{\cellcolor{white!25}{32.47\%}}\\
\hline
\textbf{Char Vectors (URLNET)}&CNN&{\cellcolor{white!25}{0.942}}&{\cellcolor{white!25}{97.08\%}}&{\cellcolor{white!25}{97.08\%}}&{\cellcolor{white!25}{2.28\%}}&{\cellcolor{white!25}{3.55\%}}&{\cellcolor{white!25}{0.006}}&{\cellcolor{white!25}{64.69\%}}&{\cellcolor{white!25}{51.54\%}}&{\cellcolor{white!25}{61.84\%}}&{\cellcolor{white!25}{35.09\%}}\\
\hline
\textbf{Word Vectors (URLNET)}&CNN&{\cellcolor{white!25}{0.956}}&{\cellcolor{white!25}{97.81\%}}&{\cellcolor{white!25}{97.81\%}}&{\cellcolor{white!25}{1.78\%}}&{\cellcolor{white!25}{2.60\%}}&{\cellcolor{white!25}{0.047}}&{\cellcolor{white!25}{61.46\%}}&{\cellcolor{white!25}{62.87\%}}&{\cellcolor{white!25}{35.69\%}}&{\cellcolor{white!25}{38.57\%}}\\
\hline
\textbf{Char + Word vectors  (URLNET)}&CNN&{\cellcolor{white!25}{0.966}}&{\cellcolor{white!25}{98.27\%}}&{\cellcolor{white!25}{98.27\%}}&{\cellcolor{white!25}{1.36\%}}&{\cellcolor{white!25}{2.10\%}}&{\cellcolor{white!25}{0.036}}&{\cellcolor{white!25}{57.98\%}}&{\cellcolor{white!25}{60.07\%}}&{\cellcolor{white!25}{37.81\%}}&{\cellcolor{white!25}{42.05\%}}\\
\hline
\textbf{Char-level+ Word Vectors  (URLNET)}&CNN&{\cellcolor{white!25}{0.955}}&{\cellcolor{white!25}{97.74\%}}&{\cellcolor{white!25}{97.74\%}}&{\cellcolor{white!25}{1.67\%}}&{\cellcolor{white!25}{2.84\%}}&{\cellcolor{white!25}{0.043}}&{\cellcolor{white!25}{61.26\%}}&{\cellcolor{white!25}{61.72\%}}&{\cellcolor{white!25}{37.81\%}}&{\cellcolor{white!25}{38.75\%}}\\
\hline
\textbf{Char + Char-level Word ( (URLNET)}&CNN&{\cellcolor{white!25}{0.960}}&{\cellcolor{white!25}{98.00\%}}&{\cellcolor{white!25}{98.00\%}}&{\cellcolor{white!25}{1.80\%}}&{\cellcolor{white!25}{2.21\%}}&{\cellcolor{white!25}{0.035}}&{\cellcolor{white!25}{57.07\%}}&{\cellcolor{white!25}{59.61\%}}&{\cellcolor{white!25}{37.81\%}}&{\cellcolor{white!25}{42.97\%}}\\
\hline
\textbf{Char Vectors (EXPOSE)}&Bag of CNN&{\cellcolor{white!25}{\textbf{0.969}}}&{\cellcolor{white!25}{98.46\%}}&{\cellcolor{white!25}{98.46\%}}&{\cellcolor{white!25}{1.50\%}}&{\cellcolor{white!25}{1.59\%}}&{\cellcolor{white!25}{0.028}}&{\cellcolor{white!25}{62.92\%}}&{\cellcolor{white!25}{57.47\%}}&{\cellcolor{white!25}{48.06\%}}&{\cellcolor{white!25}{37.00\%}}\\
\hline
\textbf{Char Vectors}&LSTM&{\cellcolor{white!25}{0.962}}&{\cellcolor{white!25}{98.1\%}}&{\cellcolor{white!25}{98.1\%}}&{\cellcolor{white!25}{1.66\%}}&{\cellcolor{white!25}{2.14\%}}&{\cellcolor{white!25}{0.035}}&{\cellcolor{white!25}{63.98\%}}&{\cellcolor{white!25}{59.41\%}}&{\cellcolor{white!25}{45.23\%}}&{\cellcolor{white!25}{35.94\%}}\\
\hline
\textbf{Char+ Word Vectors}&LSTM&{\cellcolor{white!25}{0.895}}&{\cellcolor{white!25}{94.74\%}}&{\cellcolor{white!25}{94.74\%}}&{\cellcolor{white!25}{4.54\%}}&{\cellcolor{white!25}{5.98\%}}&{\cellcolor{white!25}{0.021}}&{\cellcolor{white!25}{61.95\%}}&{\cellcolor{white!25}{55.76\%}}&{\cellcolor{white!25}{50.53\%}}&{\cellcolor{white!25}{37.95\%}}\\
\end{tabular}
 \end{small}
    \end{table*}
\begin{figure*}[!tb]
\centering{\includegraphics[width=0.75\textwidth]{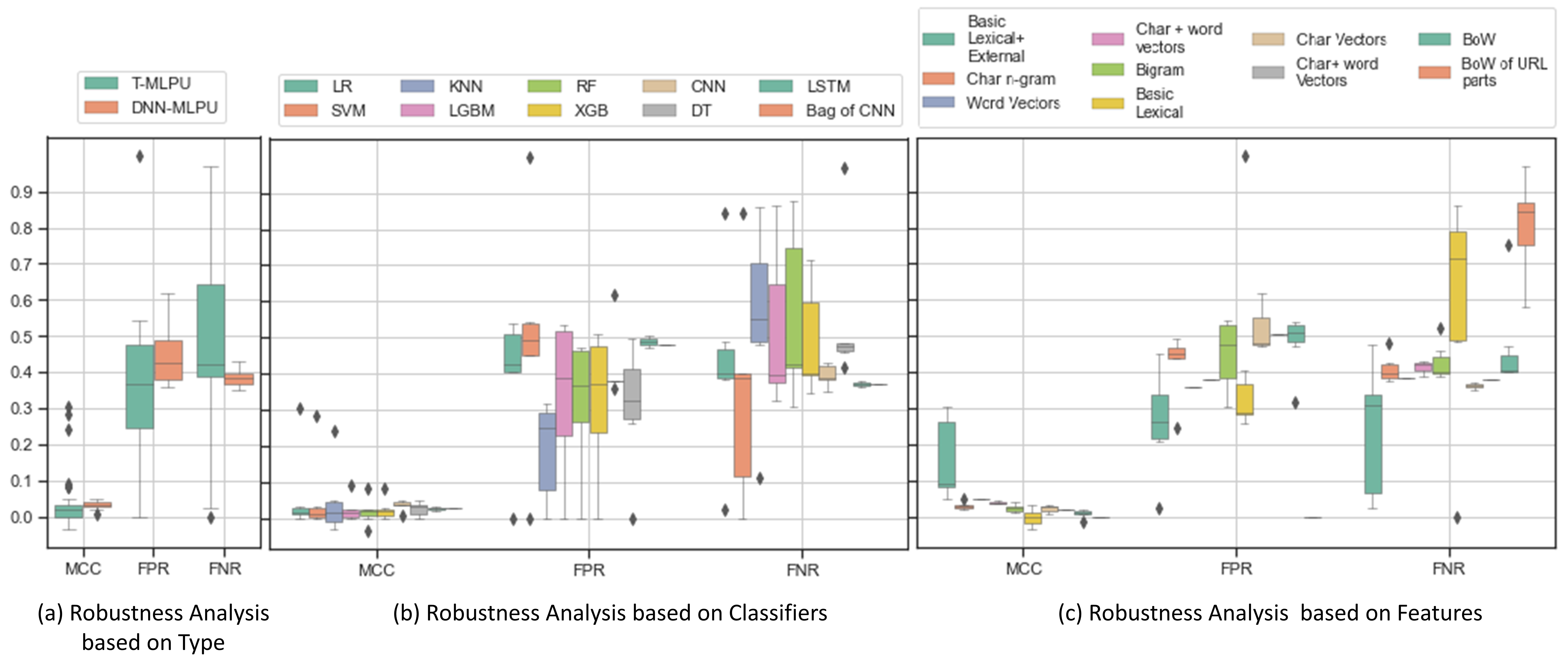}}
\centering\caption{Robustness Analysis of MLPU models}
\label{RobustnessAnalysis}
\end{figure*}
\par
\textbf{(i) Evaluate Performance (RQ2).}
\label{robustness}
Our $Adv_\mathrm{data}$ was imbalance; therefore, we evaluated robustness of  MLPU models using MCC, accuracy (ACC), AUC, FPR and FNR. 
\inserted{We used these measures as phishing URL detection can be classified as text classification problem and in NLP domain, Robustness is measured using performance on adversarial samples (please see \cite{wang2022measure} for comparison between robustness measures used in NLP versus Computer vision domain). }

\textbf{(ii) Evaluate Impact (RQ3).}
We have statistically analysed the \textit{FNR} i.e., the rate of misclassifying an AU as benign to identify the tendency of each adversary type and obfuscation method to mislead an MLPU model.
For this analysis, we have examined $10,000$ URLs of each adversary type considered boxplots and heatmaps illustration to identify the most threatening adversary type and obfuscation method.
\section{Results}
\label{results}
\subsection{URLBUG Realizability Analysis (RQ1)}
Our approach generated 1742242 valid adversarial URLs against the popular seed dataset. These URLs consisted of 34441 unique hostnames.

\textbf{(i) Computational Effort (RQ1.1).} We found that 
our method requires 23 milliseconds (ms) on average to generate  adversarial URLs for a given target URL. 
The total number of URLs $N$ generated for the domain is dependent on each adversary type i.e.,  $N \propto (n*d_f)/p$, $N \propto n*d_f*l*e$ and  $N \propto (n*d_f*t)/p$ for domain, path and TLD adversary respectively.
Here, $n$ depicts the number of URL obfuscation techniques for an adversary type, e.g., n=14,2,1 for a domain, path and TLD adversary. $l$, $e$, $t$ represent the length of path, size of extension and TLD respectively.
$d_f$ is the number of web pages in the domain containing the form HTML tag; in the best-case scenario, $d_f=$ the total number of web pages in a domain.
$p$ depicts the popularity of website brands in the phishing community, e.g., according to open-Phish \cite{OpenPhish80:online}, Pay-pal, Rediff, Apple Inc, and Bank of America are the most popular phishing target.
We found that $N$ is inversely proportional to $p$ because, for popular phishing targets, some of the generated deceptive domains were unavailable for registration, e.g., for 'Netflix'. We found that 347/848  of the generated domain names were not available for registration.

\textbf{(ii) Annual Registration Cost (RQ1.2).}
We found that 87\% of the generated adversarial URLs hostnames were available for registration using GoDaddy API \cite{httpsdev96:online} while 13\% were already taken.
Overall, 85\% out of 87\% of the generated URLs were available for registration at a median annual cost of \$11.99 and only 2\% cost more than \$11.99.
Among these, the domain and path adversary had an annual cost of maximum \$11.99, median and minimum \$8.99 while,
TLD had an annual cost of maximum  \$19999.99, median \$14.99 and minimum. 
We found that generated hostname `manage.cloud' for `manage.com' with TLD `.cloud' had a maximum price of \$19999.99/ year whereas 
most TLD adversaries using "fun, club, website, xyz, site and space TLDs are available at the lowest registration cost of \$0.99 only.
For example, the hostname` centralbankofindia.club' was only available for \$0.99. 
Hence, a famous domain name with these TLDs can quickly become a target of cybercrimes with less than a dollar of investment.
Although, there are some exceptions, e.g., `office. site' was available at \$ 6499.99 while "steampowered.site" at \$169.
Moreover, we found that adversaries with ".love", ".global" and ".host" TLDs were auctioned for most popular target such as "Netflix" and "chase"  at high prices such as "apple.host" and "chase.host" are available at a price of \$ 6499.99 and \$3249.99 respectively.

\textbf{(iii) Deception (RQ1.3).}
Despite these exceptions, 85\% out of 87\% of generated URLs were available for registration at a median annual cost of \$11.99 and only 2\% cost more than \$11.99.
Figure~\ref{AnalysisRQ1} shows the results of this analysis; we have labelled the hostnames that do not meet any criteria of deceptive URL criteria (Table~\ref{RQ1matrix}) as unknown.
Among 13\% of the AUs that were not available for registration, 61.86\% of hosts returned an HTTP status code of 200, indicating the request was successful. 
Among these, 18.35\% of hostnames did not fulfil our deceptive URL criteria.
In contrast, 43.51\% of these hostnames were deceptive, i.e., 17.5\% were legally disputed, whereas 7.5\% were intentionally protected while 18\% were redirected to another URL.
For the remaining hostname, 26.53\% of the requests resulted in an exception, suggesting that these URLs were no longer available.
For 11.5\% of these hostnames, no information was available while 
14.93\% of these fulfilled our deception criteria.
4.5\% of them were legally disputed, 9.5\% were intentionally protected and 1.8\% were redirected. Furthermore, 4.67\% URLs returned an HTTP status code of 403, showing that these are inaccessible resources among these 1.45\% fulfiled our deceptive URL criteria while no information was available for other 3.22\% URLs.
Lastly, 3.53\%, 2.46\% and 0.93\% hostnames returned an error HTTP status code of 503 (Unavailable resource error), 404 (not Found) and 406 (No acceptable response), among these 2.79\%, 0.86\% and 0.40\% were deceptive while for remaining 0.74\%, 1.60\% and 0.53\% no information was available respectively.
\begin{tcolorbox}[title=Remark 1]
URLBUG is cost-effective, as it generated URLs (requiring 23ms) that can be registered for a median annual price of \$11.99 and 81.5\% of already registered AUs were either deceptive or had unknown status.
 \end{tcolorbox}

\subsection{Robustness Evaluation (RQ2)}
\label{originalperformance}
\label{Q1}
Table~\ref{RQ1Table} shows the original performance and the adversarial performance of the considered MLPU models. The Cell is highlighted Green to show good performance i.e., for MCC, ACC, AUC, FPR, FNR of $>=0.90, >=90\%,>=90\%, <=10\%, <=10\%$ respectively, Red color depict a poor performance i.e., MCC, ACC, AUC, FPR, FNR  of $<0,<50\%,<50\%,>30\%,>30\%$. Lastly, blue color indicates otherwise.
We also highlighted MCC $=0$ with an orange colour to depict that model classify equivalent to a random guess.
\subsubsection{Original Performance}
The results showed that all the models performed well except for models trained over BoW of URL part in the original performance.
For models trained over Basic Lexical and Basic Lexical+External features, base classifiers (DT, SVM, LR and KNN) had a low performance compared to ensemble classifiers (RF, XGB and LGBM). 
Overall, among all the models, XGB trained over Char n-gram features performed the best, achieving an average MCC score of 0.972, ACC and AUC of 98.59\% and 99.86\% respectively, with the lowest FPR and FNR of 1.34\% and 1.41\%. 
Subsequently, the deep learning model Expose \cite{Saxe2017Expose} attained the second-best average MCC score of 96.91\%, ACC and AUC of 98.46\% and FPR and FNR of 1.50\% and 1.59\% respectively.
Moreover, LGBM trained over Char n-gram features also performed comparable to Expose with an average MCC of 96.63\%, accuracy and AUC of 98.31\% and 99.84\% respectively and FPR of 1.48\% and FNR 1.69\% respectively.
Lastly,  a general observation was that all the models, on average, had 1.45\% more FNR than FPR, which suggests that these models are more likely to misclassify phishing URLs as compared to benign URLs. 
\subsubsection{Overall Robustness Analysis}
\label{overallrobustness}
The results of robustness evaluation are shown by Adversarial Performance in Table~\ref{RQ1Table}. It can be seen that all the models yielded unable results against our $Adv_\mathrm{data}$.
Interestingly, the three initially best-performing models XGB and LGBM trained over Char n-gram features and Expose, were not the most robust models; instead, SVM, LR and KNN classifiers trained over Basic Lexical + External features performed better than all the other models by attaining an MCC of 0.281, 0.302, 0.241 respectively.
Moreover, SVM trained over Basic Lexical features offered more resistance to our generated AUs, attaining an FNR of 0.00\%, but it had an MCC and FPR of 0.0 and 100\%, indicating that this model classified all the URLs (including the benign seed URLs) as phishing.
Furthermore, SVM and LR trained over Basic Lexical+External features also showed resilience against our generated AUs, attaining an FNR of 2.43\% and 2.42\% and comparatively better MCC of 0.281, 0.302 than other models, respectively. However, they also misclassified 45.23\% and 40.99\% of the seed URL as phishing as depicted by their FPR.
On the contrary, DT trained over BoW of URL parts (original MCC 87.39\%) yielded the worst performance on the generated adversarial URLs with an FNR of 96.85\%.
However, this model had an FPR and MCC of zero; indeed, all classifiers trained on BoW of URL parts had MCC and FPR of zero, specifying that these features can detect benign URLs correctly they fail to identify their deceptive variants. 
Hence, we can conclude that despite few models (SVM trained over Basic Lexical and SVM, LR trained over Basic Lexical+ External features) offered resistance to our $Adv_\mathrm{data}$, their inability to detect the benign URLs correctly makes them unreliable for practical usage.

\textbf{(i) Robustness analysis based on Model Type (RQ2.1).}
To identify which type (T-MLPU vs DNN-MLPU) of MLPU models, classifiers and features offered more robustness, we performed a statistical analysis using box-plot illustration as shown in Figure~\ref{RobustnessAnalysis}. As depicted by Figure~\ref{RobustnessAnalysis} (a), DNN-MLPU models obtained a slight better median MCC and FNR equal to 0.032 and 0.382, respectively, whereas T-MLPU attained a median MCC and FNR of 0.017, 0.420 respectively.
However, DNN-MLPU had 0.060 more median FPR than T-MLPU models.
These results suggest that both model types yield unreliable outputs and fail to classify $Adv_\mathrm{data}$ correctly. 

\textbf{(ii) Robustness analysis based on classifier (RQ2.2).}
As illustrated by Figure~\ref{RobustnessAnalysis} (b), all the classifiers yielded undesirable outcomes by attaining a median MCC of 0.020.
Among these, CNN was comparatively more robust than other classifiers attaining a maximum median MCC, FPR and FNR of 0.036, 0.378 and 0.387, respectively. 
On the other hand, Bag of CNN offered more resistance to our generated AUs by attaining a minimum median FNR of 0.37.
Conversely, KNN had the lowest median FPR of 0.250, suggesting that KNN can correctly classify benign seed URLs; however, its high median FNR (0.551) depicts its inability to distinguish phishing URLs.
Moreover, LSTM misclassified benign seed URLs as phishing 11.1\% more than CNN, whereas CNN incorrectly classified our adversarial example as benign 1.6\% more than LSTM. 
Among T-MLPU classifiers, LGBM attained a comparable performance to CNN in terms of median FPR and FNR of 0.387 and
0.395 respectively. 
In summary, all the classifiers attained either high (more than 38\%) FPR or FNR.
 \begin{figure*}[!tb]
\centering{\includegraphics[width=0.85\textwidth]{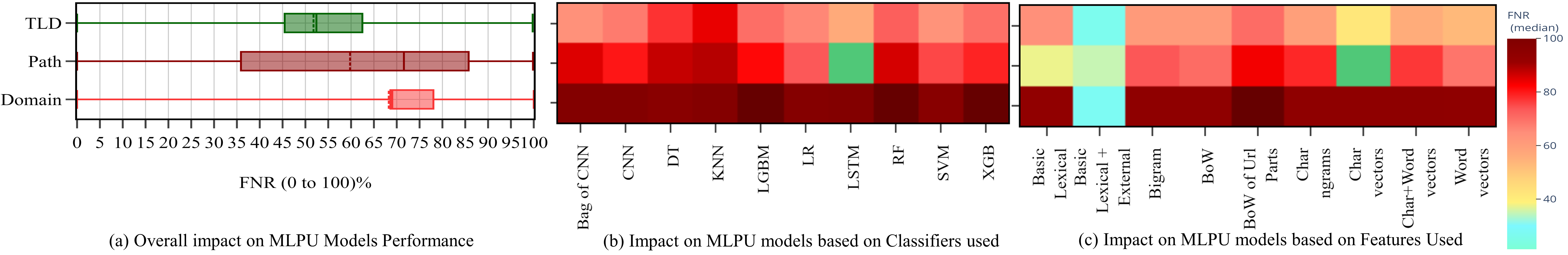}}
\centering\caption{Robustness Analysis based on Adversary Type}
\label{figure4a}
\end{figure*}
\begin{figure*}[!htb]
\centering{\includegraphics[width=0.85\textwidth]{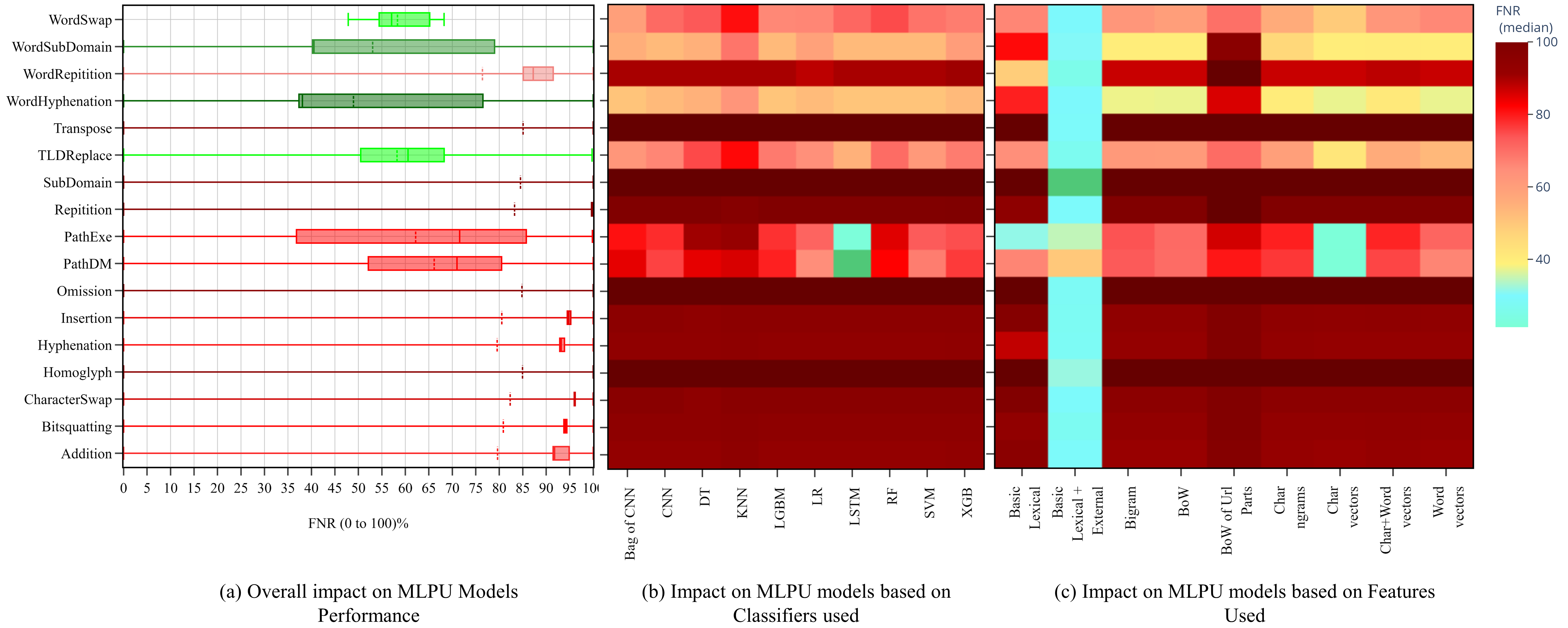}}
\centering\caption{Robustness Analysis based on URL Obfuscation Method}
\label{figure4c}
\end{figure*}
\textbf{(iii) Robustness based on Features (RQ2.3).}
We also analyzed the relationship of features used by MLPU models with their robustness against our generated AU dataset as shown by Figure~\ref{RobustnessAnalysis}c.
 Basic Lexical+ External Features
 were more resilient against our generated AUs dataset having a median MCC, FPR and FNR, i.e., 0.088,0.2615 and 0.3077
respectively) compared to other features.
These features yielded a median MCC of 5.24\% more than the best original performing char and word vector features. 
 Word vectors were the second-most robust features that attained a median MCC, FPR and FNR of 0.0472, 0.3569 and 0.3857, respectively.
Unexpectedly, char-based features (char n-gram or char vectors) combined with word vectors or alone were not robust against our $Adv_\mathrm{data}$, yielding a median MCC, FPR and FNR of 0.268, 0.4664 and 0.3868.
Lastly, Basic Lexical and BOW of URL parts attained the lowest median MCC of 0\%, indicating that these features are highly vulnerable to adversarial perturbations and cannot classify URLs correctly.
To summarize, we can conclude that Basic Lexical + External features were the most robust feature, and they were able to identify 69.22\% of our generated AUs as phishing and 73.85\% of their seeds URLs as benign.
\par
\begin{tcolorbox}[title=Remark 2]
The results imply that both traditional and deep learning MLPU models are unreliable for practical usage. However, comparatively, LR trained over Basic Lexical and External features offered more robustness for our $Adv_\mathrm{data}$ obtaining an MCC of 0.302. In terms of classifiers and features, overall CNN and Basic Lexical + External feature deemed more reliable results than others, achieving a median MCC of 0.03 and 0.08, respectively.
\end{tcolorbox}

\subsection{Robustness Analysis based on Adversary Type (RQ3)}
\label{adversarytype}
Figure~\ref{figure4a} shows the box-plot and heatmaps to illustrate the impact of each adversary type (median FNR) in terms of models, classifiers and features, respectively.

\textbf{(i) Impact of Adversary Type (RQ3.1).}
Figure~\ref{figure4a}(a) shows that URLs generated using domain and path adversary had a median FNR of 71.57\% and 68.81\%, respectively, across all models implying that MLPU models are vulnerable to these adversaries more than TLD adversaries. 
On the other hand, TLD adversary is not significantly successful, attaining a median success rate of 52.41\%. 
A reason behind this result may be that we used an already known list of malicious TLD to create these adversaries.
Thus there is a high chance that these TLDs are already present in the phishing class distribution of the training dataset.
Despite their presence, these adversaries can still deceive baseline models half of the time. 
Another interesting observation is that all adversaries have an outlier with a minimum FNR of 0\%, indicating that one of the models (SVM-Basic Lexical) can classify all the adversaries correctly.
However, we have already seen that this model had an FPR of 100\% (see section~\ref{Q1}), which means that this model classifies the original benign URLs used to generate the adversaries as phishing, thus, positing that this model is unreliable for practical usage.

Figure~\ref{figure4a}(b) shows the median FNR of classifiers against each adversary type.
The analysis reveals that all the classifiers misclassified domain adversaries with a median FNR of more than 92\%.
It suggests that domain adversaries are the hardest to detect using MLPU classifiers.
On the other hand, adversarial URLs generated using path adversaries had a variable impact on different classifiers. 
For example, we found that the LSTM classifier could detect path adversaries with 99.99\% accuracy (median FNR of 0.008\%). 
We can infer that LSTM tendency to capture the contextual information enabled it to detect path adversaries as the context (surrounding keywords order) was altered while generating these URLs (section~\ref{pathadversary}).
Furthermore, linear classifiers, SVM and LR, attained a median FNR of 66\% and 63.94\%, respectively, whereas other classifiers failed to identify path adversaries as phishing with a median FNR of $>70\%$.
Lastly, most classifiers also showed variable robustness against TLD adversaries ranging from LSTM achieving the lowest median FNR of 42.90\% to KNN attaining a highest median FNR of 76.90\%.
Additionally, SVM and Bag of CNN (Expose) were able to detect more than 48\% of TLD adversaries accurately having a median FNR of 51.43\% and 52.36\% respectively.

Figure\ref{figure4a}(c) illustrates the median FNR of the models with the features used by them. 
The results assert that Basic Lexical + External features are more robust to detect domain and TLD adversaries than other features attaining a minimum median FNR of 27.94\% and 26.80\%, respectively. 
In contrast, Char vectors are more resilient against path adversaries attaining a minimum median FNR of 21.75\%.
We noted that only external features could detect the domain adversaries (median FNR of 27.94\%), whereas other features achieved a median FNR $>90\%$. 
One rationale is that these features capture the domain's registration information irrespective of the URL's linguistic structure.
Thus, we conclude that lexical features alone are not suitable to detect domain adversaries. 
However, char vectors were also able to detect 57.88\% of the generated URLs for TLD adversaries.

\textbf{(ii) Impact of Obfuscation Method (RQ3.2).}
\label{obfuscationmethod}
Figure~\ref{figure4c} illustrates the impact of the URL obfuscation technique on the robustness of MLPU models.
Figure~\ref{figure4c}(a) shows the overall analysis. 
We reveal that the MLPU models were unable (with a median FNR of 100\%) to detect char-level URL obfuscation methods: omission, subdomain, transpose, and homoglyph. 
Besides that, these models are also susceptible (having a median FNR of $>90\%$) to other char-level URL obfuscation techniques: addition, repetition, hyphenation, character swap and insertion.
In contrast, among word-based obfuscation methods, the models performed poorly on WordRepetition  (median  FNR 87.23\%), PathExe (median FNR 71\%) and PathDM (median FNR 71\%).
Whereas, these models showed comparatively more robustness for TLDReplace, WordHyphenation and WordSubdomain based obfuscation methods, attaining a median FNR 60.57\%,  38.08\% and 40.53\% respectively.

Figure~\ref{figure4c}(b) display the analysis of classifiers performance on each URL obfuscation technique. The results reveal that the LSTM classifier was able to classify word-level URL obfuscation techniques: PathDM, PathExe, TLDReplace and WordSubDomain better than other classifiers, acquiring a median FNR of 0\%, 1.098\%, 42.90\% and 40.23\% respectively. 
In contrast, Bag of CNN (Expose \cite{Saxe2017Expose}) showed slightly more robustness (however, still insufficient) to char-level URL obfuscation techniques: Addition, Bitsquatting and Hyphenation as compared to other classifiers obtaining a median of 91.7\%, 93.78\% and 92.88 \% respectively.
LGBM detected the WordHyphenation method more effectively than other classifiers with an FNR of 37.35\% whereas, DT performed slightly better than other classifiers for detecting CharacterSwap	and Insertion methods achieving a median FNR of 93.76\% and 93.33\%, respectively.
Lastly, KNN, XGB and LR classifiers performed worst than other classifiers in detecting obfuscated URLs. 
These results imply that only LSTM and LGBM classifiers showed appropriate robustness for PathDM, PathExe and WordHyphenation methods.

Figure~\ref{figure4c}(c) exhibit the analysis of URL obfuscation methods with the features used by the models.
Basic Lexical + External are more robust (with a median FNR of 28.46\%) against all the URL obfuscation methods except for Pathexe and PathDM than other features.
For Pathexe and PathDM, we found that char vectors outperformed all the other features by obtaining a median FNR of 21.67\% and 21.83\%, respectively.
BoW of URL parts performed the worst among all the features to detect all the URL obfuscation techniques attaining a median FNR of 99.37\%.
WordSubDomain and Homoglyph techniques are the hardest to be detected by all the features, even the most robust feature; Basic Lexical + External had a median FNR of 30\%.

\begin{tcolorbox}[title=Remark 3]
LSTM classifier can detect path and TLD adversaries more accurately, while among features, Basic Lexical+ External are more robust against the domain and TLD adversaries, while char-vectors can detect path adversaries more accurately.
Moreover, we found that MLPU systems are susceptible to character-level perturbations. Among classifiers, LSTM showed more robustness to detect PathDM and PathExe. Basic Lexical +External are comparatively reliable features achieving a low ($<$28\%) FNR  across all the URL obfuscation techniques except for PathDM and PathExe. 
\end{tcolorbox}

\section{Discussion and Analysis}
\label{discussion}
\subsection{Security Vulnerabilities}
\label{Vulnerabilites}
We found the following security vulnerabilities in the considered MLPU detectors by analysing our study's results.

\textbf{(i) Incomplete Modelling Assumption.}
The MLPU models learn the decision boundary based on the lexical or syntactic differences between phishing and legitimate URL tokens. 
For example, the classifiers trained over Basic Lexical features assume that there exists a distinguishable difference between the structure of legitimate versus phishing URLs
e.g., the length of the URLs.
Similarly, n-gram and vector-based models assume that character and word distribution between both classes is separable. 
However, these models did not consider the case where there is a strong correlation between the legitimate and phishing URL.
For example, for legitimate URL \url{hxxps://www.paypal.com} and phishing URL against it \url{hxxps://paypal.com.igr2.ru/n}, 
there is a significant similarity instead of difference. 
Our adversarial URLs exploited this vulnerability and generated deceptive URLs with lexical, syntactic, and semantic similarity with their targets (legitimate URLs). Consequently, our results have demonstrated that our generated AEs successfully evaded the considered MLPU models.
\textit{Therefore, we assert that researchers should investigate new techniques that can extend the coverage of the MLPU models to capture both the similarity and difference aspects between legitimate and phishing URLs. 
For example, combining current MLPU models with techniques from other domains such as Doc2Vec (that captures the similarity between different documents in text classification tasks) \cite{lau2016empirical} may extend the coverage of MLPU systems beyond capturing only differences. 
We believe such techniques may result in more robust MLPU models.}

 \begin{figure}[t]
 \centering
\includegraphics[width=\columnwidth]{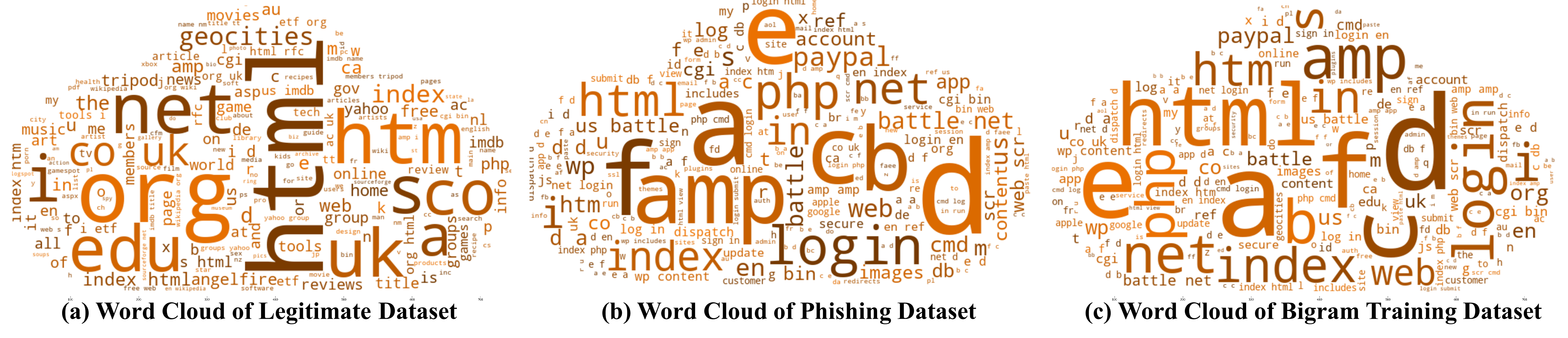}
\caption{Distribution of Phishing and Legitimate Vocabulary}
\label{fig:sub-second}
\end{figure}

\textbf{(ii) Bias towards Popular Phishing Targets.}
Our results (section~\ref{overallrobustness}) have revealed that all models attained a high FPR for the benign seed URLs.
We speculate a bias in the decision boundary of traditional and deep learning MLPU systems, i.e., legitimate URLs of popular and potential phishing targets are misclassified as phishing.
On further analysis, we found that this issue was limited to domain names and extended to other recurring tokens in the phishing dataset. 
MLPU systems misclassified all the examples with login keywords irrespective of a URL's legitimate or phishing nature.
e.g., \url{hxxps://www.netflix.com/login} was misclassified by all models.
It implies that 
these models cannot distinguish between phishing and legitimate URLs due to the inclination towards phishing tokens.
This limitation is partially visible in the FPR (median 2.53\%) of the models' initial validation results (Table~\ref{RQ1Table}).
However, it is less evident because the models are trained over the large dataset, and the FPR is not often significantly large.
We believe the reason behind the high FPR can be that for a
given website, the domain name can be repeated in the training dataset at most to the number of web pages on a website 
whereas the probability of the same path, query-string and parameter repetition across all legitimate URLs is relatively low.
Subsequently, for phishing URLs, 
an attacker usually generates multiple URLs of the same kind \cite{Saxe2017Expose} 
targeting popular brands. 
Therefore, the probability of phishing tokens appearing as features is high but limited to specific brands.
 We further investigated our assumption by analyzing the BoW vocabulary of legitimate and phishing datasets individually and comparing it with the combined training dataset vocabulary. Figure~\ref{fig:sub-second} illustrates this analysis. 
We found that most of the words appearing in the BoW features of the training dataset are high-frequency words from phishing URLs. 
 In contrast, only a few domain names from the legitimate dataset (Geocities and Wikipedia) with more web pages appear in the vocabulary.
 It suggests that the training vocabulary is more biased towards phishing tokens while legitimate domain names remain underrepresented. Attackers can exploit this vulnerability by generating AUs against less popular targets and with more legitimate keywords as also shown by our results.
 \textit{We hope that the research community investigates the relationship of FPR of their MLPU models with phishing dataset trends and provide solutions to mitigate this vulnerability for developing robust MLPU models.}

\textbf{(iii) Limited Obfuscation Methods in Phishing Dataset.}
Phishing URL datasets collected by the current studies are obtained from blacklists such as PhishTank, Openphish during a specific time frame. 
Our results (section~\ref{realizabilityevaluation}) reveal that valid and realizable deceptive URLs can be generated easily.
The ease of generating and registering these deceptive URLs poses a real threat to the MLPU models.
It suggests that MLPU models trained on original Phishing datasets do not represent the Phishing URLs' universal pattern; hence, MLPU models are evaded successfully. 
\textit{We propose that researchers conduct more studies to identify, generate and evaluate diverse adversarial URLs for MLPU models and augment them with the current datasets to enhance the robustness of MLPU models.} 
 \begin{figure}[!tb]
\centering
\includegraphics[width=\columnwidth]{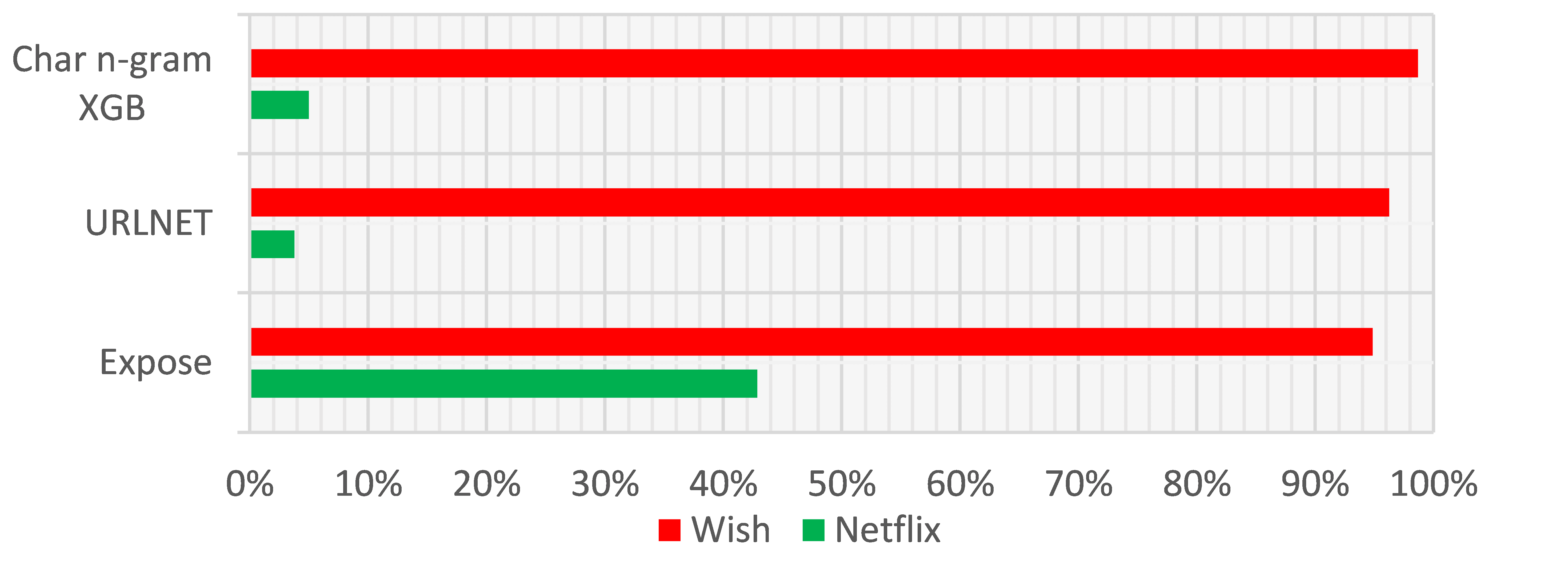}
\caption{Biases towards Phishing Target Trend}
\label{fig:sub-first}
\end{figure}

\subsection{ Defense against Adversarial URLs} 
\inserted{Our results revealed that the MLPU models are unreliable.
Yet, we didn't find any study done to protect these MLPU models against evasion attacks. Therefore, we explored whether state-of-the-art defences can enhance the robustness of MLPU models. For this, we applied two popular defences used in Adversarial Machine learning literature to defend against evasion attacks: (i) Adversarial Training \cite{shrivastava2017learning} and (ii) Ensemble Learning \cite{shu2022omni,li2020adversarial}.}

\begin{figure*}[!tb]
\centering{\includegraphics[width=0.9\textwidth]{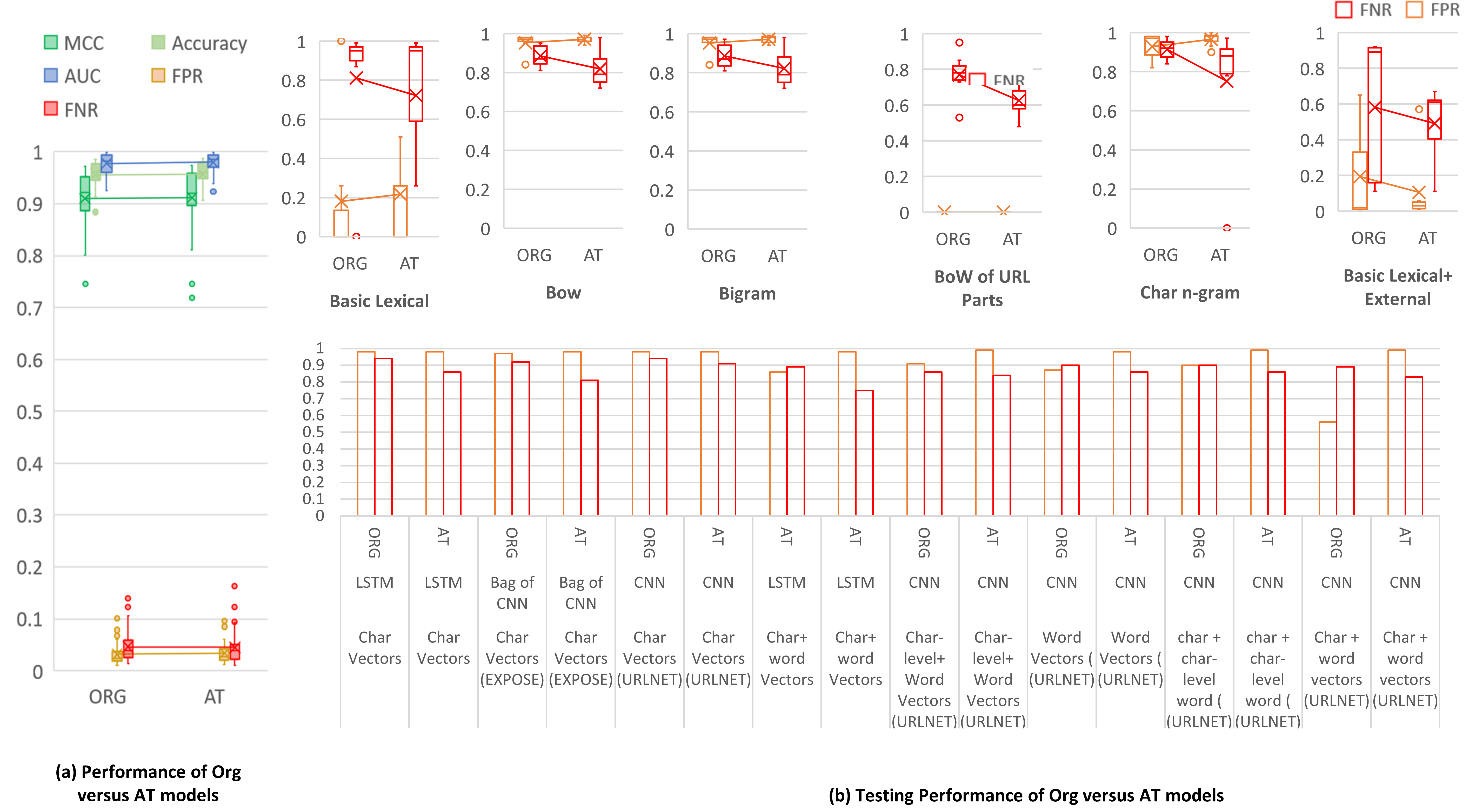}}
\centering\caption{Original (ORG) versus Adversarial Trained (AT) Models Performance Analysis}
\label{RQ4Table}
\end{figure*}
\par
\inserted{\textbf{(i) Adversarial Training (AT).}
Adversarial training is a practice used in 
image \cite{shrivastava2017learning} and text \cite{ren2019generating} classification to improve model robustness ML-based systems. 
}

\inserted{\textit{Training.} 
For adversarial training, we used a similar setting as reported by the studies \cite{jin2019bert, li2018textbugger} and concatenated the adversarial samples generated using the S dataset to perform adversarial training.
We hyper-tuned all the models using the same settings as mentioned in section \ref{hypertuning}.
Lastly, we recorded the performance of adversarially trained models using the same measures as discussed in section \ref{origperf}.
Figure~\ref{RQ4Table}(a) depicts the statistical analysis of the performance of original versus adversarially trained models using boxplot illustration. We observed that AT insignificantly impacts the cross-validation performance of original MLPU models which implies that AT preserves the information of original URLs and also captures our augmented Adversarial URL knowledge. 
For testing the AT models on a novel dataset, we acquired a new dataset S' and generated adversarial examples against it. 
The dataset was attained from Similar web \cite{SimilarW15} and yahoo finance \cite{UNVEILED49} and consisted of a list of top e-commerce and shopping sites in Australia. 
We removed URLs present in the S dataset (popular phishing targets) to avoid biases in testing results as adversarial training was performed using deceptive URLs generated from S. 
Finally, we ended up with 107 e-commerce site URLs absent in the $S$ seed URL dataset.
Although these URLs were not popular phishing targets, they represented potential threats as they require the user's personal information such as credit cards to make a purchase.
Some examples of URLs in S' are woolworth.com, wish.com, ozbargain.com, JBHIFI.com, alibaba.com, and target.com. The complete list of these URLs is made available online \cite{RobustEVALMLPU} for research validity.
}
\inserted{
Figure~\ref{RQ4Table}(b) illustrates the results of this comparison. Overall AT T-MLPU models performed better than the original model in detecting new deceptive URLs as indicated by the decrease in FNR. In contrast, original models were better than AT-trained models in detecting legitimate URLs as benign (as shown by an increase in FPR of AT models).
\textit{Interestingly, AT models trained over Basic Lexical + External features performed better than their original counterparts as evident by their median FPR and FNR. 
On the other side, DNN-MLPU performance insignificantly changed after Adversarial training. Instead, the FPR of the models increased after AT. }}

\inserted{\textbf{(ii) Ensemble Learning.}
Ensemble learning is another popular method adapted by researchers to improve the robustness of models. In this approach, multiple classifiers are combined to improve classifier robustness \cite{shu2022omni}. 
Based on our results in section ~\ref{overallrobustness}, we selected models trained on Basic Lexical + External features and the LSTM model for creating a meta-classifier to achieve resilience. Then we trained the Logistic Regression meta-classifier by stacking the selected models. 
Our results indicated that the model performed perfectly as we attained a 10-fold cross-validation performance of 100\% for MCC, ACC, AUC, recall, precision and 0\% FPR and FNR. }

\inserted{We then tested the meta-classifier against Adversarial URLs generated for popular and future phishing targets datasets (S and S') respectively. For popular adversarial URLs, we attained better robustness results than original MLPU models with an MCC of 0.088, ACC of 0.67, AUC of 0.73, FPR of 0.20 and 0.32.
However, for future adversarial URLs, the meta-classifier still returned unreliable results i.e., MCC of 0.009, ACC of 0.08, AUC of 0.53, FPR of 0 and FNR of 0.92. 
\textit{These results suggest that the state-of-the-art adversarial defences on MLPU models can only help to protect popular phishing targets against URLBUG. 
Moreover, ensembling MLPU models with other types of phishing detectors such as visual similarity (VSB) and web-content (WCB) based can further improve the phishing detection paradigm but these solutions will also be limited to popular phishing targets (section~\ref{introduction}). 
We recommend that more studies should be conducted to design scalable and robust phishing detectors that can also handle attacks on less popular phishing targets. }
}
\subsection{Evaluation Challenges}
\label{challenges}
\textbf{(i) Lack of Reproducible MLPU models:}
While conducting our study, we found that the current research on MLPU models lack reproducibility. This subsection discusses reproducibility challenges faced by the current state-of-the-art MLPU research and provides researchers with guidelines to fill this gap.
\textit{a) Availability of Datasets}: 
Datasets used by most of the prior studies \cite{kolari2006svms, urlnet, Saxe2017Expose} are not publicly available and therefore reproducing their work is not possible.
For instance, a study \cite{urlnet} was trained on a dataset on 5M records, but it was collected privately from VirusTotal \cite{virustotal} and was not made public.
Another observation is that several studies \cite{huang2012svm,Saxe2017Expose} utilized online threat intelligence sources such as PhishTank \cite{PhishTan46:online} and OpenPhish \cite{OpenPhish80:online} to collect the dataset for training their models.
These feeds are time-dependent, i.e., phishing URLs present in these sources in 2016 might be obsolete in 2021,
However, most of the studies did not mention the time and criteria used to collect the dataset, making it difficult to reproduce the prior works.
We propose that the researchers should make their dataset public and provide documentation for it (such as time criteria used to collect the dataset) to enhance research reproducibility. 
A recent survey made a similar observation on ML-based approaches for data exfiltration detection \cite{sabir2020machine}.
\textit{b) Availability of Code and Trained Model}
The absence of a reproducible code is one of the primary challenges faced for replicating the prior works. 
More specifically, none of the studies Traditional ML-based phishing URLs detectors (Table~\ref{stateofart}) made their code available for research validity and re-usage. 
We recommend that future research focus on reproducibility and help the research community validate and adversarial evaluate their work.
\textit{c) Diversity of Performance Metrics and Evaluation}
We observed that the prior works used two metrics frequently to report their work performance, i.e., accuracy and AUC (Table~\ref{stateofart}).
However, several of the previous studies in other domains asserted that accuracy is not a reliable measure to evaluate a model's performance, especially when models are trained on an imbalance dataset \cite{luque2019impact,chicco2020advantages}.
Likewise, we observed (Table~\ref{RQ1Table}) that for traditional MLPU models, AUC overestimates the performance of a model that suffers from the high FNR.
Therefore, accuracy and AUC alone are not suitable measures for evaluating the performance of these models. 
\textit{We suggest that future research propose a standard evaluation benchmark to evaluate the performance of MLPU models to assist their comparison and enrichment.}
\begin{table*}[!htb]
\caption{\inserted{Additional Analysis - Popular Character-level attacks on MLPU models [Valid refers to structural validity of the generated URL according to RFC format, while Adversarial indicates that the new URL is valid and has deceptive host-name]}}
\label{additionalattacks}
\begin{small}
\centering
\resizebox{\textwidth}{!}{\begin{tabular}{l|l|l|l|lll}
\hline
Attack&Model&Skipped&Failed&\multicolumn{3}{|c}{Successful}\\\hline
&&&&Total&Valid &Adversarial\\
\hline
DeepWordBug \cite{Gao2018DeepWordBug}&	\multirow{3}{*}{EXPOSE (Bag of CNN)}&	78.82\%&	0.00\%&	21.18\%&	7.49\%&	1.02\%\\
Robust Word Recognition \cite{pruthi2019combating}	&&	78.82\%&	0.28\%&	19.52\%&	7.49\%&	1.94\%\\
PWWS \cite{ren2019generating}	&&	78.82\%&	2.31\%&	18.87\%&	8.23\%&	0.00\%\\
\hline
DeepWordBug\cite{Gao2018DeepWordBug}&\multirow{3}{*}{Char Vector LSTM}	&77.24\%&	3.70\%&	19.06\%&	5.92\%&	2.31\%\\
Robust Word Recognition\cite{pruthi2019combating}&	&77.24\%&	\textbf{9.99}\%&	12.77\%&	6.66\%&	3.79\%\\
PWWS\cite{ren2019generating}&		&77.24\%&	9.34\%&	13.41\%&	6.11\%&	0.65\%\\
\hline
DeepWordBug	&\multirow{3}{*}{Char + Word vectors (URLNET)}&	72.90\%&	0.00\%&	\textbf{27.10}\%&	\textbf{16.00}\%&	0.37\%\\
Robust Word Recognition	&&	72.90\%&	2.68\%&	24.42\%&	11.93\%&	2.68\%\\
PWWS	&&72.90\%&	1.02\%&	26.09\%&	11.93\%&	2.68\%\\
\hline
DeepWordBug	&\multirow{3}{*}{Char + Char-level Word (URLNET)}	&77.98\%&	0.00\%&	22.02\%&	13.88\%&	1.39\%\\
Robust Word Recognition	&&	77.98\%&	3.70\%&	18.32\%&	10.64\%&	\textbf{5.00}\%\\
PWWS	&&77.98\%&	0.74\%&	21.28\%&	15.26\%&	0.28\%\\

\hline
\end{tabular}}
\end{small}
\end{table*}

\textbf{(ii) Lack of Adversarial Evaluation.}
Although few works have been done on evaluating the security of MLPU models \cite{AlEroud2020GAN, DeepPhis17, AdversarialSampling}, they are limited to specific MLPU models and cannot be extended to other models. Similarly, they have not publicly published the AUs, hindering research reproducibility. We encourage researchers to make their AUs available to evaluate MLPU models. As a first step, we have made our $Adv_\mathrm{data}$ available \cite{RobustEVALMLPU}. This dataset can educate the users about URL obfuscation techniques that they should consider while deciding to click a suspicious URL. Moreover, this dataset can be augmented with the original training data to enhance the robustness of MLPU models (Adversarial Training). \textit{Lastly, we assert that developers of MLPU models should not merely rely on the performance of the models but also assess their model's robustness and reliability by identifying evasion scenarios.}

\inserted{\textbf{(iii) Lack of popular adversarial attack transferability.}
While conducting our study, we considered augmenting popular black-box char-level word substitution adversarial attacks such as DeepWordBug \cite{Gao2018DeepWordBug}, Robust Word Recognition \cite{pruthi2019combating} and PWWS \cite{ren2019generating} using TextAttack library \cite{morris2020textattack}.
We considered four top (MCC$>$0.96) DNN-MLPU models (Table \ref{RQ1Table}) and used 1080 popular seed URLs to generate adversarial examples using these attacks. The results of this analysis are shown in Table~\ref{additionalattacks}.
The analysis reveals that most of the attacks are skipped due to the high FPR of these models on popular seed URLs (see section~\ref{Vulnerabilites} (iii)). Moreover, we discovered that Char Vector LSTM offered more resilience to these attacks as compared to the other top-performing DNN-MLPU models as illustrated by failed attempts. Furthermore, among the successful attacks, DeepWordBug outperformed all the other attacks, yet, only a few of these successful attacks resulted in URLs with valid structure \cite{RFC3986U96:online}.  Lastly, as these attacks are originally developed for text classification systems, these attacks didn't result in considerable adversarial URLs, with maximum (5\%) adversarial URLs generated by Robust Word Recognition \cite{pang2019improving} attack against Char + Char-level Word (URLNET). These outcomes exhibit that NLP character-level attacks are somehow successful but are inadequate to generate deceptive URLs.}

\inserted{Likewise, we also assessed two word-substitution-based attacks from NLP: TextFooler \cite{jin2019bert} and TextBugger \cite{li2018textbugger}. Both of the attacks employed the synonym substitution technique to spawn AEs and utilised semantic and grammatical similarity to evaluate the realizability of generated AEs automatically. 
However, both of these attacks were unsuccessful in generating valid deceptive URLs. For example, TextFooler \cite{jin2019bert} generated `hxxp://hxxp.yelp.ra' for the seed URL `hxxp://www.google.ae'. This URL is not deceptive as it doesn't target the Google brand name. Similarly, the attack returned MySpace as Facebook synonym and Gmail as Office synonym. This implies that synonym-based word substitution attacks don't generate deceptive adversarial URL variants. One of the reasons behind it is that, unlike text, a URL is a \textit{sequence of characters} instead of meaningful English words, and synonym substitution is not possible in most cases.
\textit{From these results, we endorse that more efforts should be done on developing novel methods for generating adversarial URLs to enrich the robustness of evaluation datasets.}
}

\inserted{\textbf{(iv) Lack of Robustness Metrics. }}
\inserted{
Robustness Verification provides a guarantee that the model prediction does not change on small perceivable perturbations.
In the text classification domain, robustness is primarily measured using performance on adversarial samples \cite{wang2022measure}. 
However, lately, several efforts have been done by the research community to formally verify the robustness of Deep Neural Networks (DNN) \cite{ma2018deepgauge,dong2020empirical,zhang2020machine}. 
Nonetheless, most of the endeavours are done in the computer vision domain and the produced metrics are not directly relevant to a constrained discrete domain like URL classification \cite{Zhang2019TextualAdversarialExamplesSurvey,rosenberg2021adversarial}.
For illustration, CLEVER (Cross Lipschitz Extreme Value for nEtwork Robustness) \cite{weng2018evaluating} score is designed to compute `robustness lower bounds' for gradient-obfuscated DNN under the white-box setting. This approach is only appropriate to process continuous gradient-based perturbations on image data and classifiers. Whereas, gradient-based perturbations are not directly applicable to symbolic textual data   \cite{wang2022measure, Zhang2019TextualAdversarialExamplesSurvey} 
}.
\inserted{Recently few works in NLP domains have presented formal robustness verification metrics such as Interval Bounded Propagation (IBP) \cite{huang2019achieving}, Maximal Safe Radius (MSR) \cite{la2020assessing}, Auto-LIPRA \cite{xu2020automatic} and semantic robustness \cite{la2022king}. Though most of them only consider word-substitution-based perturbations in white-box settings and are not directly appropriate for MLPU detectors. For instance, the MSR method computes the upper and lower bounds for the text DNN using synonym-based perturbations. However, synonym-based perturbations are not suitable for deceptive URL classification task (see section \ref{challenges} (iii)). Finally, we studied the IBP technique to compute the robustness of NLP models using symbol substitution, however, we concluded that the IBP method is used to train verifiable robust models and is not used for computing the robustness of pre-trained DNN models in black-box settings. 
\textit{From this assessment of the state-of-the-art, we propound that there is a need to extend robustness evaluation metrics beyond computer vision and the NLP domain. Moreover, we highlight that robustness metrics should be extended from white-box to black-box settings to avoid retraining the models for evaluating robustness bounds. Eventually, We stress that future research is needed for developing robustness metrics for phishing detectors to ensure their reliability.} }
\subsection{Limitation of our study and Future Work}
 (i) The source codes and datasets for traditional ML models proposed by prior studies were unavailable. Consequently, our experimental setup may not precisely replicate these baseline models. However, to mitigate this threat, we tried to replicate the MLPU systems (classifiers and features) reported by the previous studies and fine-tuned them to select optimal MLPU models. We further tried to use large vocabulary sizes for n-gram features to keep our deployment aligned with these baseline models and reported their performance using multiple state-of-the-art performance measures. (ii) {We restricted the scope of the word-level mutations (except TLD adversary) to English words. Although, one may argue that the top-phished brands have a global presence. A rationale behind limiting these adversaries to English words was to increase the understandability and verify them. However, this is one of the limitations of our study}.
(iii) Due to the unavailability of large public phishing datasets, data collection threat exists in our training dataset. To mitigate this threat, we collected a large dataset obtained from various sources (used by the prior studies such as OpenPhish \cite{OpenPhish80:online}, PhishTank \cite{PhishTan46:online}).
(iv)  Due to a lack of research transparency among the MLPU community, we can not generalize our findings to an industry or open-source solution using different features and classifiers. However, to encourage research transparency, validity and reproducibility, we have made our code, trained models and dataset available \footnote{\url{https://figshare.com/s/45bc4e48eec57e849afd}}. 
(v) In this study, we tested the robustness of MLPU systems only.
\textit{In the future, we plan to investigate the robustness of ML-based phishing detectors based on WCB and VSB systems to get a complete picture of the state-of-the-art ML-based phishing detectors. Moreover, we plan to develop a robust and reliable MLPU model by using the insights gained from this study. Lastly, we intend to extend NLP robustness verification measures (as discussion in section~\ref{discussion} to phishing domain i.e., ``How to develop formally certified roboustness methods for MLPU detectors under different perturbations?''}

\section{Conclusion}
\label{conclude}
Machine learning-based Phishing URL detectors (MLPU) have been extensively proposed since the past decade to classify URLs as phishing in a timely and scalable manner.
However, the robustness of these models against adversarial manipulation remains comparatively unknown. 
In this study, we have proposed a methodology to test the robustness of MLPU models. Our work extensively and comprehensively benchmarks the robustness of 50 state-of-the-art MLPU models by considering both traditional Machine and Deep learning models. 
Our results and analysis have identified reliable MLPU models, classifiers, features, and potent adversaries. Furthermore, it has unveiled several security vulnerabilities and evaluation challenges of these models. Some of our key findings are: (i) our generated AUs are realizable (valid URLs and can be registered at a median annual price of \$11.99) and deceptive (63.94\% of the registered URLs are used  malicious purposes).
(ii) High performing (MCC$>$ 0.90) MLPU models yield unreliable (MCC $<$ 0.31) results when subjected to AUs.
(iii) Basic Lexical+ External and character vector features offer more resilient (misclassify $<$30\%) to adversarial test cases than other features.
(iv) LSTM classifier has more tendency (accuracy 99.99\%) to detect path adversary.
(v)  MLPU models are susceptible (misclassify 90\%) to character-level perturbed URLs.
We hope this study is received as a call to action to address novel security vulnerabilities and challenges in developing robust MLPU models.

\bibliographystyle{IEEEtran}
\bibliography{reference.bib}
\vskip -2\baselineskip plus -1fil
\begin{IEEEbiography}[{\includegraphics[width=1in,height=1.25in,clip,keepaspectratio]{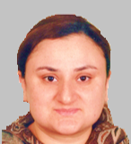}}]{Bushra Sabir}
Bushra Sabir is currently a PhD student in the school of computer science, University of Adelaide and CSIRO's 
Data61. Her area of research is adversarial Machine Learning for cybersecurity systems. She is especially interested in the area of Secure Machine learning for detecting data exfiltration attacks such as Phishing, Malicious Domains, Spam text and Insider threat. Bushra utilized Machine and Deep learning methods to create adversarial examples of cybersecurity applications (such as phishing URLs, spam emails). In her PhD, she will expand her research works to explore solutions to handle adversarial evasion attacks on Machine and Deep learning cybersecurity systems. Previously she has worked as a lecturer for six years and worked in the software industry for two years. She has achieved a gold medal in Master of Computer and Software Engineering from National University of Technology (NUST-Pakistan).
\end{IEEEbiography}
\vskip -2\baselineskip plus -1fil
\begin{IEEEbiography}[{\includegraphics[width=1in,height=1.25in,clip,keepaspectratio]{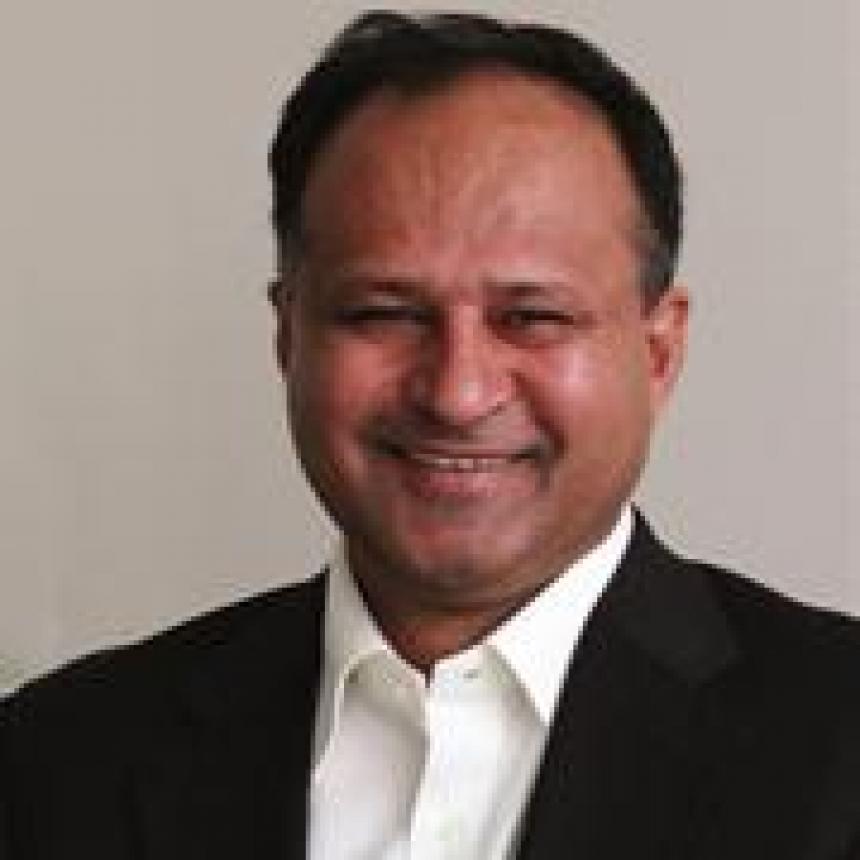}}]{M. Ali Babar}
M Ali Babar is a Professor in the School of Computer Science, University of Adelaide. He is an honorary visiting
professor at the Software Institute, Nanjing University, China. He has authored/co-authored more than 230 peerreviewed
papers in premier Software Technology journals and conferences. With an H-Index 48, the level of
citations to his publications is among the leading Software Engineering researchers in Aus/NZ. At the University
of Adelaide, Professor Babar has established an interdisciplinary research centre, CREST-Centre for Research on
Engineering Software Technologies, where he leads the research and research training of more than 30 members.
He has been involved in attracting several millions of dollar worth of research resources over the last ten years.
Prof Babar leads the University of Adelaide’s participation in the Cyber Security Cooperative Research Centre
(CSCRC), one of the largest Cyber Security initiative of the Australasian region. Within the CSCRC, he leads the
theme on Platform and Architecture for Cyber Security as a Service. Further details can be found on:
\url{https://researchers.adelaide.edu.au/profile/ali.babar#my-research}.
\end{IEEEbiography}
\vskip -2\baselineskip plus -1fil
\begin{IEEEbiography}[{\includegraphics[width=1in,height=1.25in,clip,keepaspectratio]{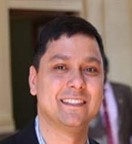}}]{Dr Raj Gaire}
Raj Gaire is a Senior Experimental Scientist at CSIRO
Data61. He has been leading cross-domain collaborations
for data driven innovation and has been actively contributing to Agriculture, Energy and Healthcare domains. He
has worked as a software architect, analyst and developer in CSIRO and other organisations for over 15 years.
His research interests are Distributed Computing, Cloud
Computing, IoT, Semantic Web, Cyber Security and Data
Analytics including Bioinformatics, Social Media Analytics
and Big Data. He received his Ph.D. from the University of
Melbourne, Australia
\end{IEEEbiography}
\vskip -2\baselineskip plus -1fil
\begin{IEEEbiography}[{\includegraphics[width=1in,height=1.25in,clip,keepaspectratio]{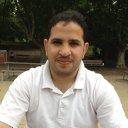}}]{Dr Alsharif Abuadbba}
Dr Alsharif  Abuadbba is a Senior Research Scientist at CSIRO
Data61. He has a PhD in computer security from RMIT university, Melbourne, Australia 2017. He was the winner of RMIT prestigious VC award of Research Excellence in Technology 2018. He was also the winner of Eureka innovation award 2018 for his cybersecurity startup named EyeCura. Dr Alsharif is a Senior Research Scientist at CSIRO's Data61.
Dr Abuadbba has previously worked with Californian based technology companies, OptCTS Inc and AgilePQ Inc, as a senior R\&D engineer and contributed to a few US IP patents in cybersecurity. He has mentored 30+ interns and joiner engineers. He also has several publications in high-quality venues. Dr Abuadbba has recently helped in defining Data61 Cyber Security CRC projects, such as Deception as a Service and Smart Shield which got +\$2M dollar funds.  His specialist and interests include AI for cybersecurity, security of AI, IoT, cryptography and watermarking.
\end{IEEEbiography}
\end{document}